\documentclass[a4paper,11pt]{article}
\usepackage[margin=1in]{geometry}
\usepackage{epsfig}
\usepackage{cite}
\usepackage{graphicx}
\usepackage{epstopdf}
\usepackage{color}
\UseRawInputEncoding
\usepackage[nottoc]{tocbibind}
\usepackage[english]{babel}
\usepackage{tabularx}
\usepackage{xparse}
\usepackage{amsmath}
\usepackage{hyperref}
\usepackage{grfext}
\usepackage{amssymb}
\usepackage{caption}
\usepackage{mathtools}
\usepackage{subcaption}
\usepackage{algorithm2e}
\usepackage{float}
\usepackage{subcaption}
\usepackage{multirow}
\restylefloat{table}
\usepackage{booktabs}
\usepackage[dvipsnames]{xcolor}
\usepackage{tikz}
\usepackage{pdflscape}
\usepackage{lscape}

\usepackage{amsmath}
\usepackage{amssymb}
\usepackage{caption}
\usepackage{mathtools}
\usepackage{graphicx}
\usepackage{hyperref}
\usepackage{subcaption}
\usepackage{algorithm2e}
\usepackage{float}
\usepackage[nameinlink]{cleveref}
\usepackage[toc,page]{appendix}
\usepackage{float}
\usepackage{subcaption}
\usepackage{multirow}
\restylefloat{table}
\usepackage{booktabs}
\usepackage[dvipsnames]{xcolor}
\usepackage{tikz}
\usepackage{pdflscape}
\usepackage{comment}

\date{\today}
 \normalsize

\newcommand{\bmat}{\left(\begin{array}}
\newcommand{\emat}{\end{array}\right)}
\newcommand{\be}{\begin{equation}}
\newcommand{\ee}{\end{equation}}
\newcommand{\bea}{\begin{eqnarray}}
\newcommand{\eea}{\end{eqnarray}}

\def\gtwid{\mathrel{\raise.3ex\hbox{$>$\kern-.75em\lower1ex\hbox{$\sim$}}}}
\def\ltwid{\mathrel{\raise.3ex\hbox{$<$\kern-.75em\lower1ex\hbox{$\sim$}}}}
\def\gev{{\rm \, Ge\kern-0.125em V}}
\def\tev{{\rm \, Te\kern-0.125em V}}

\def    \be            {\begin{equation}}
\def    \ee            {\end{equation}}
\def    \bea           {\begin{eqnarray}}
\def    \eea           {\end{eqnarray}}

\def\a{\alpha}

\def\g{\gamma}

\def\m{\mu}
\def\nn{\nonumber}

\begin{document}
\renewcommand{\thefootnote}{\fnsymbol{footnote}}

\vspace{.3cm}

\title{\Large\bf Dynamics and stability of the two-body problem with Yukawa correction to Newton's gravity, revisited and applied numerically to the solar system}

\author
{ \hspace{-3.cm} \it \bf  Nawras Abo Hasan$^{1}$\thanks{seagull18990@gmail.com}, Nabil Joudieh$^{1}$\thanks{njoudieh@yahoo.fr} and Nidal Chamoun$^{2}$\thanks{nidal.chamoun@hiast.edu.sy}
 \\\hspace{-3.cm}
 \footnotesize$^1$ Physics Department, Damascus University, Damascus, Syria \\\hspace{-3.cm}
\footnotesize$^2$  Physics Department, HIAST, P.O. Box 31983, Damascus, Syria \\\hspace{-3.cm}
}

\date{}
%\date{\today}

\maketitle
%\begin{center}
%\small{\bf Abstract}\\[3mm]
%\end{center}
\begin{abstract}
In this manuscript, we review the motion of two-body celestial system (planet-sun) for a Yukawa-type correction on Newton's gravitational potential using Hamilton's formulation. We reexamine the stability using the corresponding linearization Jacobian matrix, and verify that the Bertrand's theorem conditions are met for radii $\ll 10^{15} m$, and so bound closed orbits are expected.
Applied to the solar system, we present the equation of motion of the planet, then solve it both analytically and numerically. Making use of the analytical expression of the orbit, we estimate the Yukawa strength $\a$, and find it larger than the nominal value ($10^{-8}$) adopted in previous studies, in that it is of order ($\a = 10^{-4}-10^{-5}$) for terrestrial planets (Mercury, Venus, earth, Mars and Pluto) whereas it is even larger ($\a = 10^{-3}$) for the Giant planets (Jupiter, Saturn, Uranus and Neptune).
Taking as inputs ($r_{min}, v_{mas}, e$) observed by NASA, we analyse the orbits analytically and numerically for both the estimated and nominal values of $\a$, and determine the corresponding trajectories. For each obtained orbit we recalculate the characterizing parameters ($r_{min}, r_{max}, a, b, e $) and compare their values according to the used potential (Newton with/without Yukawa correction) and to the method used (analytical and/or numerical).
When compared to the observational data, we conclude that the correction on the path due to Yukawa correction is of order of and up to 80 million km (20 million km)  as a maximum deviation occurring for Neptune (Pluto) for nominal (estimated) value of $\a$.
\end{abstract}

\maketitle


\vskip 0.3cm \hrule \vskip 0.5cm'
%\newpage
%**************************************************************
%**************************************************************

%\begin{document}
%%%%%%%%%%%%%%%%%%%%%%%%%%%%%%%%%%%%%%%%%%
\setcounter{section}{-1} %% Remove this when starting to work on the template.
%\section{How to Use this Template}

\section{Introduction}
The past several years have witnessed a resurgence of interest in experimental testing of gravity, particularly in the possibility of deviations from the predictions of Newtonian gravity, which is considered as an excellent approximation of General Relativity (GR) on large distance scale \cite{0}. Many theoretical models suggest the existence of new, relatively weak, intermediate-range force coexisting with gravity such that the net resulting interaction would behave like a new correction to the potentials defining the gravitational field. It is known \cite{1} that there are only two types of central potentials, namely the Newton $1⁄r$ and the Harmonic $r^2$ potentials, where ANY finite motion of an object, subject to this central potential, leads to a closed path (Bertrand’s theorem). There are some `exceptions' to this statement, in the sense that there might be closed bound trajectories for a central potential different from the Newton and Harmonic ones, which have been studied in \cite{2,3}. In this contribution, we revisit the effect of a Yukawa correction to the gravitational force over large distances.

Theories of massive gravity \cite{Hinterbichler,deRham}, adding a mass term to the graviton (the carrier of gravity), have raised a wide interest and the Yukawa potential is the popular parametrization of such theories. Actually, many works describing deviations from Newton’s inverse square law have addressed the Yukawa-type correction. Assuming gravity is exerted by exchanges of gravitons, it is clear that a test for the graviton mass ($\m_g$) is to ask whether the Newton ($1/r$) potential shows any evidence of dying at large distances because of Yukawa exponential cutoff ($e^{-\m_g r}$).  Since the seventies \cite{Goldhaber}, bounds on the gravitons mass ($\m_g \leq 1.1 \times 10^{-29}$ eV) were used to put a bound on its compton wavelength considered as a distance scale for Yukawa correction ($\lambda \sim \frac{2\pi}{\m_g} \geq 3.7$ MPc). The authors of \cite{Dong} gave a bound on the Yukawa range ($\lambda$) in the order of ($10^1 - 10^4$ AU) corresponding to ($\m_g \leq 10^{-24}$) eV.

 Theories like Scalar-Tensor-Vector Gravity Theory \cite{4} predict a Yukawa-like fifth force. The authors of \cite{Zhang}, showed that screened modified gravity can suppress the fifth force in dense regions and allow theories to evade the solar system and laboratory tests of the weak equivalence principle.  In \cite{DAddio}, an extended theory of  gravity, with a modified potential including post-Newtonian terms, whose expansion is different from that of Yukawa correction, called `vacuum bootstrapped Newtonian gravity', was subjected to solar system tests, through a procedure which was applied to Yukawa corrections at the Galactic center \cite{Monica}, with no significant deviations from GR found.

 In \cite{DAMTP}, a Keplerian-type parametrization was shown as a solution of the  equations of motion for a Yukawa-type potential between two bodies. In fact, the two-body solution for alternative theories yield a strong constraint for solar system \cite{Banik,Lu}, whereas several analyses of Yukawa potential for a 2-body system in different contexts were carried out \cite{Pricopi,Edwards}. The orbit of a single particle moving under Yukawa potential was studied in \cite{Mukherjee}, and the precessing ellipse type orbits were observed. In \cite{Iorio}, it was noted that the modified gravity with Yukawa-like long-range potential was (un)successful on astrophysical scales (in solar system), whereas an analysis of Yukawa potential in $f(R)$ gravity was given in \cite{Laurentis}. The work of \cite{Berge} showed that a Yukawa fifth force is expected to be
sub-dominant in satellite dynamics and space geodesy experiments, as long as they are
performed at altitudes greater than a few hundred kilometres. The Yukawa strength was estimated in \cite{Lorenzo} to be ($\a < 10^{-5}$-$10^{-8}$) for distances of order $10^9$ cm, whereas the use of laser data from LAGEOS satellites yield a constraint on $\a$ of the order of $10^{-12}$.

In this letter, we build on work from \cite{5}, in which  the dynamics and stability of the two body problem with a Newtonian potential corrected by a Yukawa term were explored. In particular, we reproduced their analytical results and applied them to the study of all the planets of our solar system.
Solving analytically the planet equation of motion, one finds an elliptical trajectory, which one can also obtain numerically using Runge-Kutta method. Starting from the observed values of the perihelion distance and velocity $(r_{min}, v_{max})$ and of the tranjectory eccentricity $e$, stated in NASA public results \cite{6}\footnote{Although the  standard deviations of the planetary trajectories are not quoted in the NASA public website, however one can consider that the corresponding error equals to the last digit of the quoted significant numbers. In our computations, we used the whole digits allowed by machine precision, however the results in the appendices tables showed only significant digits equal to those of the observed data.}, one could determine the ellipsis equation and estimate, for Yukawa corrected potential, the Yukawa strength $\a$. One can use this estimated value, or another nominal value taken from other studies, to either draw the analytical trajectory and recalculate the characterizing parameters: the shortest (longest) distance to the Sun $r_{min}(r_{max})$, the semi major (minor) axis $a(b)$ and the eccentricity $e$, or to solve numerically the equations of motion with the Yukawa-corrected potential in order to check
the closedness of the resulting trajectory, whose characteristics are to be reevaluated again.
 Later, we compared these results with those calculated for the Keplerian motion of planets subject to the pure Newtonian potential, and,  in addition, showed the compatibility of the results with the observational NASA data.

More specifically, for the two-body system (planet-sun), the Newtonian potential is given by:
\begin{equation} \label{eq1}
V_N(r)=-\frac{G m_p M_{\odot}}{r}
\end{equation}
where $G=6.674 \times 10^{-11} \frac{\mathrm{Nm}^2}{\mathrm{Kg}^2}$ is the gravitational Newton constant, $m_p$ ($M_\odot$) is the planet (sun) mass. With a Yukawa correction, the gravitational potential becomes
\begin{equation}\label{eq2}
V(r)=-\frac{G m_p M_{\odot}}{r}\left(1+\alpha e^{-\frac{r}{\lambda}}\right)=V_N(r)+V_{Y k}(r)
\end{equation}
where $V_{Yk}$ is the Yukawa correction to the Newtonian potential and $\alpha$ ($\lambda$) represents the strength (range) of the Yukawa correction. Previous studies \cite{5,7} gave the nominal values ($\a= 10^{-8} (\lambda = 10^3 \mbox{AU}=10^{15} \mbox{m})$. However, our estimations gave a larger order of magnitude for the Yukawa strength: $\a \sim 10^{-4}-10^{-5}$ for terrestrial planets (Mercury, Venus, Earth, Mars and Pluto) and $\a \sim 10^{-3}$ for the remaining Giant planets (Jupiter, Saturn, Uranus and Neptune), which are in line with \cite{DAMTP}.

We saw that for estimated $\a$, the maximum deviation from observed data, which increases the further the planet is (20 million km in Pluto), is less than that of the $\a$ nominal value (80 million km in Neptune), which is plausible considering that the estimation of $\a$ is done by identifying the factor containing it to observational data.

For each of the nominal and estimated values of $\a$, we analysed the planet's trajectory both analytically and numerically. Analytics wise, we started from the observational data of NASA ($r_{min}, v_{max}, e$) and reconstructed the closed ellipse trajectory of which we re-evaluated the characteristics ($r_{min}, r_{max}, a, b, e$) and compared with the pure Newton case and with the observational data. Numerics wise, the $\a$ determines the potential under which the planet moves, and so one can solve the equations of motion numerically using Runge-Kutta method taking as initial conditions the observed data of ($r_{min}, v_{max}$), to check that one gets  closed trajectories in excellent agreement with the elliptical shapes, of which we can evaluate the characteristics that one compares to the pure Newtonian case, to the analytical method results and to the observed data.

The manuscript is organized as follows. In section (1), we revise the system dynamics using Hamilton's method. In section(2), we state the types of stability and determine the one corresponding to the system under study. We discuss,  in section(3) and following \cite{5}, Bertrand's theorem and get the analytical solution to the equation of motion. Finally, we apply in section (4) the obtained approximative analytical results to  the study of the solar system planets in order to estimate the Yukawa strength and re-determine the trajectory characteristics for both estimated and nominal values of $\a$,  as well as solve numerically the equations. The results, of comparing the analytical/numerical outputs with the observed data according to the used potential, are presented in form of plots for all the planets, whereas the corresponding tables are given in an appendix. We end up with conclusions in section (5).

\section{Hamiltonian formulation}
\begin{figure}[tbp]
\centering
\includegraphics[width=13cm, height=8cm]{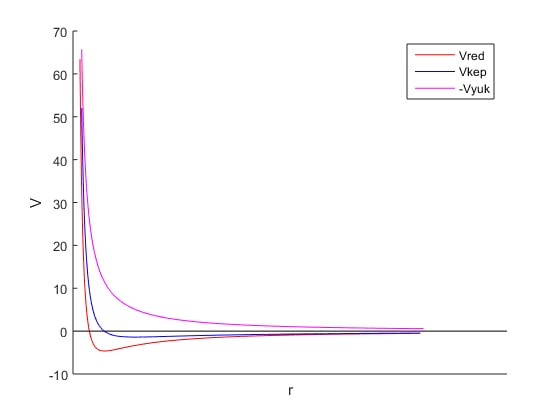}
\caption{\label{fig1}  The reduced potential (red line) given for fixed angular momentum (Eq. \ref{eq11}).
The pink line  denotes the magnitude of the purely Yukawa term $(\left|-\frac{\a K}{r} e^{-\frac{r}{\lambda}}\right|$), whereas the blue line represents the Keplerian reduced potential, i.e. Eq. \ref{eq11} without the Yukawa term.}
\end{figure}

We start with the Hamiltonian $H=T+V$ where $T$ is the kinetic energy of both masses and $V$ is the Gravitational potential energy.
\begin{equation}\label{eq3}
\mathcal{H}=\frac{\vec{p}_1^2}{2 m_p}+\frac{\vec{p}_2^2}{2 M_{\odot}}-\frac{K}{\left|\vec{r}_2-\vec{r}_1\right|}\left(1+\alpha e^{-\frac{\left|\vec{r}_2-\vec{r}_1\right|}{\lambda}}\right)
\end{equation}
where $\vec{r}_i, (\vec{v_i}), i=1,2$ are the positions (velocities) of the two masses with corresponding momenta $p_1=M_{\odot} v_1, p_2=m_p v_2$, $K=G m_p M_{\odot}$.
Changing to the center of mass frame (c.o.m), with
\begin{eqnarray}
 \vec{r}_{1}=+\frac{m_p}{m_p+M_\odot} \vec{r} = + \frac{\mu}{M_\odot} \vec{r} + \vec{R} &,& \vec{r}_{2}=-\frac{M_\odot}{m_p+M_\odot} \vec{r} = - \frac{\mu}{m_p} \vec{r} + \vec{R} \label{r1r2}\\
   \vec{r}=\vec{r}_1-\vec{r}_2 &,&  \vec{R}=\frac{M_\odot \vec{r}_1 + m_p \vec{r}_2}{m_p+M_\odot} \label{rR} \\
   \vec{v}_1 = \dot{\vec{R}} + \frac{\mu}{M_{\odot}} \vec{v} &,& \vec{v}_2 = \dot{\vec{R}} + \frac{-\mu}{m_p} \vec{v} \label{v1v2} \\
   \vec{v} = \dot{\vec{r}}&,& \vec{p}=\m \vec{v} \label{vp}, \\
     \ddot{\vec{R}}=\vec{0} &,& \mu \ddot{\vec{r}} = M_{\odot} \ddot{\vec{r}}_1 = - m_p \ddot{\vec{r}}_2 \label{ddotRr},
  \end{eqnarray}
we get
\begin{eqnarray}\label{eq4}
\mathcal{H}=
\frac{1}{2} (M_{\odot}+m_p) \dot{\vec{R}}^2 + H &:& H= \frac{p^2}{2 \m} -\frac{K}{r}\left(1+\alpha e^{-\frac{r}{\lambda}}\right)
\end{eqnarray}
Here we have defined $\mu=\frac{m_p M_{\odot}}{m_p+M_{\odot}}$ as the reduced mass of the system and $r=|\vec{r}|$. We switch to polar coordinates in the c.o.m to get
\begin{equation} \label{eq5}
H=\frac{1}{2 \mu}\left(p_r^2+\frac{p_{\varphi}^2}{r^2}\right)-\frac{K}{r}\left(1+\alpha e^{-\frac{r}{\lambda}}\right)
\end{equation}
From the canonical equations (\cite{8}):
$\dot{q}_i=\left(\frac{\partial H}{\partial p_i}\right), \dot{p}_i=-\left(\frac{\partial H}{\partial q_i}\right)$, and since the Hamiltonian is cyclic in $\varphi$ (i.e. it does not depend explicitly on $\varphi$), we have:
%\begin{equation}
\begin{align}
%\begin{gathered} \label{eq6}
\dot{\varphi}=\frac{\partial H}{\partial p_{\varphi}}=\frac{p_{\varphi}}{\mu r^2} \\ \label{eq7}
\dot{p}_{\varphi}=-\frac{\partial H}{\partial \varphi}=0 \Rightarrow p_{\varphi}=\mu r^2 \dot{\varphi}=\ell= \mbox{constant}
%\end{gathered}
\end{align}
%\end{equation}
where $\ell$ is the angular momentum of the two-body system, and therefore Hamilton’s equations for $r$ become:
%\begin{equation}
\begin{align}
%\begin{gathered}
\label{eq8}
\dot{r}=\frac{\partial H}{\partial p_r}=\frac{p_r}{\mu} \\ \label{eq9}
\dot{p}_r=-\frac{\partial H}{\partial r}=\frac{\ell^2}{\mu r^3}-\frac{K}{r^2}\left[1+\alpha\left(1+\frac{r}{\lambda}\right) e^{-\frac{r}{\lambda}}\right]
%\end{gathered}
\end{align}
%\end{equation}
Again, and since $H(t)=H(t_0)=h$ is constant during the motion of the masses \cite{8}, and since $p_r^2=\mu^2 \dot{r}^2 \geq 0$ we get a lower bound for the total energy of the system:
\begin{equation} \label{eq10}
h \geq \frac{\ell^2}{2 \mu r^2}-\frac{K}{r}\left(1+\alpha e^{-\frac{r}{\lambda}}\right)
\end{equation}
The right hand side of eq. (\ref{eq10}) is defined to be the ``reduced potential'', which is common in the Kepler problem moving from two degrees of freedom to only one (with the Yukawa correction)
\begin{equation} \label{eq11}
V_{r e d}(r)=\frac{\ell^2}{2 \mu r^2}-\frac{K}{r}\left(1+\alpha e^{-\frac{r}{\lambda}}\right)
\end{equation}
One can draw the function for fixed $\ell$ giving the allowed regions of motion (look at figure \ref{fig1}). Note that $\mu >0,\lambda >0$ and $\alpha > 0$.

\section{The linearization matrix}
Following \cite{9}, in order to determine the stability of the equilibrium points of the system, we must form a matrix differential equation using the system equations of motion (Hamilton’s Eqs. \ref{eq8} and \ref{eq9} for $r, p$). The linear system has the form:
\begin{equation} \label{eq12}
\begin{aligned}
&\frac{d}{d t}\left(\begin{array}{c}
r \\
p_r
\end{array}\right)=\left(\begin{array}{c} f(r,p)\\g(r,p)\end{array}\right)=\left(\begin{array}{c} f_0\\g_0\end{array}\right)_{eq}+\left(\begin{array}{ll}
\frac{\partial f}{\partial r} & \frac{\partial f}{\partial p_r} \\
\frac{\partial g}{\partial r} & \frac{\partial g}{\partial p_r}
\end{array}\right)\left(\begin{array}{c}
r \\
p_r
\end{array}\right)\\ \mbox{where }
&f\left(r, p_r\right)=\frac{p_r}{\mu}, \quad g\left(r, p_r\right)=\frac{\ell^2}{\mu r^3}-\frac{K}{r^2}\left[1+\alpha\left(1+\frac{r}{\lambda}\right) e^{-\frac{r}{\lambda}}\right]
\end{aligned}
\end{equation}
Given that $\lambda =10^{15}m$ for orbits of size comparable to the solar system dimensions \cite{10}, one can assume that $\frac{r}{\lambda}$ is small enough that one can Taylor expand the exponential and ignore terms of $\left( \frac{r^2}{\lambda^2} \right)$, leading to:
\begin{equation} \label{eq13}
e^{-\frac{r}{\lambda}} \approx 1-\frac{r}{\lambda}+O\left(\frac{r^2}{\lambda^2}\right)\approx 1-\frac{r}{\lambda}
\end{equation}
Thus \[g\left(r, p_r\right)=\frac{\ell^2}{\mu r^3}-\frac{K}{r^2}\left[1+\alpha
\left(1+\frac{r}{\lambda}\right)
\left(1-\frac{r}{\lambda}\right)\right] \approx
\frac{\ell^2}{\mu r^3}-\frac{K}{r^2}(1+\alpha),
\] with the Yukawa effect within this approximation being limited to replacing $K$ by $K(1+\a)$, which tells that the potential shape is still Newtonian ($1/r$), and according to Bertrand's theorem every bound trajectory is thus closed for small $r/\lambda$. One can see this fact directly from Eq. (\ref{eq11}) as it gives, compared to the Keplerian potential, within the approximation just a shift, in addition to the replacement ($K \rightarrow K(1+\a)$), which does not interfere in the equations of motion:
\begin{eqnarray}
V_{red}(r) &\approx& \frac{\ell^2}{2 \m r^2} - \frac{K}{r} (1+\a) + \frac{K\a}{\lambda}.
\end{eqnarray}
Consequently, the Jacobian matrix takes the form:
\begin{equation} \label{eq14}
\left(\begin{array}{c}
\dot{r} \\
\dot{p}_r
\end{array}\right)=\left(\begin{array}{cc}
0 & \frac{1}{\mu} \\
\frac{-3 \ell^2}{\mu r^4}+\frac{2 K}{r^3}(1+\alpha) & 0
\end{array}\right)\left(\begin{array}{c}
r \\
p_r
\end{array}\right)
\end{equation}
where terms of order $O\left( \frac{r^2}{\lambda^2} \right)$ were ignored, and where the equilibrium point $(r,p_r)_{eq}$ satisfies $f_{eq}(r,p_r)=g_{eq}(r,p_r)=0$.
We can determine the $r$ at equilibrium using (eq. \ref{eq9}) to get upto leading order:
\begin{comment}
\begin{equation} \label{eq15}
\frac{l^2}{\mu r^3}-\frac{K}{r^2}\left[1+\alpha\left(1+\frac{r}{\lambda}\right)\left(1-\frac{r}{\lambda}\right)\right]=0
\end{equation}
Again  the equilibrium points are determined by solving the following equation:
\begin{equation} \label{eq16}
\frac{l^2}{\mu}-K r(1+\alpha)=0
\end{equation}
Solving this equation gives us the equilibrium solution:
\end{comment}
\begin{equation} \label{eq17}
r_{e q}=\frac{\ell^2}{\mu K(1+\alpha)}
\end{equation}
We can now test for stability by choosing values of ($\alpha, \mu, K, \ell , \lambda$) and finding the eigenvalues of the Jacobian matrix (\ref{eq14}) after substituting the equilibrium solution found above (eq. \ref{eq17}). Recall that the eigenvalues $\beta_1, \beta_2$ are found by solving the following equation:
\begin{equation} \label{eq18}
\operatorname{det}\left|J-\beta I_{2 \times 2}\right|=0
\end{equation}
with $I_{2
 \times 2}$ referring to the $2\times 2$ identity matrix. Thus we have
\begin{equation} \label{eq19}
\left|\begin{array}{cc}
-\beta & \frac{1}{\mu} \\
\frac{-3 \ell^2}{\mu r^4}+\frac{2 K}{r^3}(1+\alpha) & -\beta
\end{array}\right|=0
\end{equation}
The characteristic equation (the eigenvalue equation) becomes:
\begin{align}
%\begin{equation}
%\begin{gathered}
\beta_{1,2}=\frac{1}{2}\left(\tau \pm \sqrt{\tau^2-4 \Delta}\right) \label{eq20}\\
\tau=\operatorname{trace}(J)=0 \label{eq21}\\
\Delta=\operatorname{det}(J)=\frac{\mu^2 K^4(1+\alpha)^4}{\ell^6} \label{eq22}
%\end{gathered}
\end{align}
%end{equation}
Following \cite{11}, the stability is determined by the sign of the eigenvalues. Since $\Delta >0$, we have the following cases:
\begin{itemize}
\item	$\tau <0 ,\tau^2-4\Delta >0 \Rightarrow (r_0,p_{r0} )$  a stable node.
\item	$\tau <0 ,\tau^2-4\Delta <0 \Rightarrow (r_0,p_{r0})$ a stable spiral.
\item	$\tau >0 ,\tau^2-4\Delta >0 \Rightarrow (r_0,p_{r0})$  an unstable node.
\item	$\tau >0 ,\tau^2-4\Delta <0 \Rightarrow (r_0,p_{r0})$ an unstable spiral.
\item	$\tau =0 ,\tau^2-4\Delta <0 \Rightarrow (r_0,p_{r0})$ a neutrally stable center (which is our case).
\end{itemize}
Actually, the stability refers to how the solution behaves near the equilibrium point; in that unstable solutions grow to infinity, whereas stable solutions tend to zero. Also, it is the imaginary cases which are the ones giving bound orbital solutions (specifically the center case, whereas the stable and unstable imaginary cases are bound solutions tending towards or away from zero).

\section{Stability \& Bertrand’s theorem}
First, we rewrite the eigenvalue equation in the form
\begin{equation} \label{eq23}
\beta^2+\frac{\mu^2 K^4(1+\alpha)^4}{\ell^6}=0
\end{equation}
leading to:
\begin{equation}\label{eq26}
\beta=\pm i \frac{\mu K^2(1+\alpha)^2}{\ell^3}
\end{equation}
We note that one can study the case for a purely Newtonian Potential by letting $\alpha \rightarrow 0$. Similarly, by ignoring the terms derived from the Newtonian potential, one can single out the pure Yukawa contribution. In these two extreme cases, the characteristic equations becomes
\begin{align}
%\begin{gathered}
\mbox{Pure Newtonian: }\beta^2+\frac{\mu^2 K^4}{\ell^6}=0 \label{eq24}\\
\mbox{Pure Yukawa: }\beta^2+\frac{\mu^2 K^4 \alpha^4}{\ell^6}=0 \label{eq25}
%\end{gathered}
\end{align}
giving
\begin{align}
%\begin{gathered}
\mbox{Pure Newtonian: }\beta=\pm i \frac{\mu K^2}{\ell^3} \label{eq27}\\
\mbox{Pure Yukawa: }\beta=\pm i \frac{\mu K^2 \alpha^2}{\ell^3} \label{eq28}
%\end{gathered}
\end{align}
Thus, the equilibrium points for the purely Newtonian, the purely Yukawa, and the Newton plus Yukawa Potentials remain center solutions. This implies that the motion would remain restricted to ellipses about the equilibrium point; and so, orbits near the equilibrium point are possible (further away from the equilibrium point one would have unbounded solutions, as Fig. \ref{fig1} shows). This proves that for small $r/\lambda$ we have stable, closed orbits.

For the Keplerian orbit equation, it can be written as:
\begin{equation} \label{eq29}
\frac{d^2 u}{d \varphi^2}+u=-\frac{\m}{\ell^2} \frac{d}{d u} V\left(\frac{1}{u}\right)
\end{equation}
where $u=1⁄r$ denotes the Binet transformation, giving, for small $r/\lambda$, the following differential equation:
\begin{equation}\label{eq33}
\frac{d^2 u}{d \varphi^2}+u=+\frac{\mu K}{\ell^2}(1+\alpha)
\end{equation}
whose solution is given by
\begin{equation}\label{eq34}
u(\varphi)=\frac{1}{r}=A\left[1+e \cos \left(\varphi-\varphi_0\right)\right]: A= \frac{\mu K}{\ell^2}(1+\alpha)
\end{equation}
with $e$ is the eccentricity of the orbit. The purely Newtonian and purely Yukawa cases follow respectively from (\ref{eq33})
%\begin{equation}
\begin{align}
\mbox{Newtonoian:  }
&u(\varphi)=\frac{1}{r}=\frac{\mu K}{\ell^2}\left[1+e \cos \left(\varphi-\varphi_0\right)\right] \label{eq35}\\
\mbox{Purely Yukawa: }
&u(\varphi)=\frac{1}{r}=\frac{\mu K \alpha}{\ell^2}\left[1+e \cos \left(\varphi-\varphi_0\right)\right] \label{eq36}
\end{align}
%\end{equation}
Finally, in order to satisfy Bertrand’s theorem, the following condition should be satisfied
\begin{equation}\label{eq37}
\left.\frac{d^2 V_{r e d}(r)}{d r^2}\right|_{r=r_0}>0
\end{equation}
where the reduced potential is given by (\ref{eq11}). With the approximations of (eq. \ref{eq13})) and ignoring terms of order $O\left( \frac{r^2}{\lambda^2} \right)$ this condition becomes
\begin{equation} \label{eq38}
\left.\frac{d^2 V_{r e d}(r)}{d r^2}\right|_{r=r_0}=\frac{\mu^2 K^4(1+\alpha)^4}{\ell^6}>0
\end{equation}
which is true, since $\alpha,\mu,K,\ell>0$, in general and in the special cases of Newtonian ($\alpha=0$) and purely Yukawa potentials. This shows that the Yukawa plus Newtonian potential satisfies Bertrand’s theorem for small $r⁄\lambda$. %A consequence of Bertrand’s theorem is that the ratio $\omega/\dot{\phi}$ is a rational number; where $\omega = 2\pi⁄T$ and $\dot{\phi}$ is given by (eq. \ref{eq6}).

\section{Application to the solar system}
We present here our results consisting of determining first the parameters of the models ($r_{min}, r_{max}, a, b, e$) by comparing the previous approximative analytical solutions with the NASA data. Then, we solved the equations of motion numerically using Matlab and the fourth-order Runge-Kutta method with no approximation so that to be compared with the analytical solutions and with the observed NASA data. We applied this for all the planets of the solar system. For each pair (sun-planet) we used the following values $M_{\odot}=1.9885 \times 10^{30} \mbox{kg},\alpha_{\mbox{nominal}} =10^{-8}, \lambda =10^{15} \mbox{m}$. We list in Table (1) the initial conditions used in the analytical and numerical calculations (the period $\tau$ is used only in the numerical solution to determine the corresponding `step'):
%\begin{landscape}

\begin{table}[h]
%\scalebox{0.8}{
\hspace{-3.cm}
\begin{tabular}{p{18mm}|l|l|l|l|l|l|l|l|l|}
\cline{1-10}
 & MERCURY & VENUS   & EARTH    & MARS    & JUPITER  & SATURN   & URANUS   & NEPTUNE & PLUTO    \\ \cline{1-10}
$m_{p}$$(\times 10^{24} kg)$ & 0.3302  & 4.8673  & 5.9722   & 0.64169 & 1898.13  & 568.32   & 86.811   & 102.409 & 0.01303  \\
\cline{1-10}
$\tau $ (days) & 87.969  & 224.701 & 365.256  & 686.98  & 4332.589 & 10832.33 & 30685.4  & 60189   & 90560    \\ \cline{1-10}
 %$r_{0}$  $(\times 10^6 km)$& 0.05791 & 0.10821 & 0.149597 & 0.22794 & 0.778479 & 1.432041 & 2.867043 & 4.5034  & 5.869656 \\ \cline{1-10}
 $r_{min}$ $(\times 10^6 km)$ & 0.046   & 0.10748 & 0.147095 & 0.20665 & 0.740595 & 1.357554 & 2.732696 & 4.47105 & 4.434987 \\ \cline{1-10}
 $v_{max}$ $(\times 10^3 m/s)$ & 58.98   & 35.26   & 30.29    & 26.5    & 13.72    & 10.18    & 7.11     & 5.5     & 6.1      \\ \cline{1-10}
 eccentricity& 0.20563 & 0.00677 & 0.01671  & 0.09341 & 0.04839  & 0.05415  & 0.04717  & 0.00859 & 0.24881  \\ \cline{1-10}
\end{tabular}
%}
\caption{Initial conditions used in the calculations where $m_{p}$ denotes the planet mass, $\tau$ is the orbit period, $r_{min}$ is the perihelion and $v_{max}$ denotes the perihelion velocity} \label{tab1}
\end{table}
%\end{landscape}

\subsection{\bf Analytical Method}
The analytical ellipsis equation is of the form
\begin{eqnarray} \label{ellipsiseq}
\frac{1}{r} \equiv u &=& \frac{a}{b^2} \left( 1+e \cos \varphi\right),
\end{eqnarray}
where (for a $y$-axis perpendicular to the polar axis in the orbit plane)
\begin{eqnarray}
r_{min}=a(1-e) &,& r_{max}=a(1+e), \label{rminrmax}\\
e=\frac{c}{a}=\sqrt{1-\frac{b^2}{a^2}} &:& c^2=a^2-b^2, \label{ec}\\
a=\frac{r_{min}+r_{max}}{2} &,& b= \frac{y_{max}-y_{min}}{2} \label{ab}
\end{eqnarray}
Thus, analytically one can start with ($r_{min}, v_{max}, e$) observed by NASA in \cite{6} to compute\footnote{Due to measurement errors and orbits not being perfectly elliptical, the NASA data may give slightly different values of $a$ using Eq. \ref{ab} or Eq. \ref{abBye}.   }:
\bea
a=\frac{r_{min}}{1-e} &,& b=a\sqrt{1-e^2}, \label{abBye}
\eea
and estimate the strength $\a$ from
\bea
\frac{\m K}{\ell^2} (1+\a) = \frac{a}{b^2} &\mbox{using}& \ell=r_{min} v_{max}.
\eea
Once the analytical equation is determined, then one can plot the trajectory and recompute the characteristics ($r_{min}, r_{max}, a , b, e$) using Eqs (\ref{rminrmax},\ref{ec}). We call this procedure the ``analytical-$\a$-estimated'' approach.

One can also use the nominal value of $\a=10^{-8}$, and plug it in Eq. (\ref{ellipsiseq}), where $\ell, e$ are taken from the observed data, to re-evaluate ($r_{min}, r_{max}, a, b, e$) from
\bea
a=\frac{1}{A (1-e^2)}, &A= \frac{\m K (1+\a)}{\ell^2}, & b=\frac{1}{A \sqrt{1-e^2}}.
\eea
We call this procedure the ``analytical-$\a$-nominal" approach, which can be looked at as a method with three inputs $(\a,\ell,e)$ instead of the three inputs ($r_{min}, v_{max}, e$) used in the other approach.

\subsection{\bf Numerical Method}
\begin{figure}[tbp]
\centering
\includegraphics[width=13cm, height=8cm]{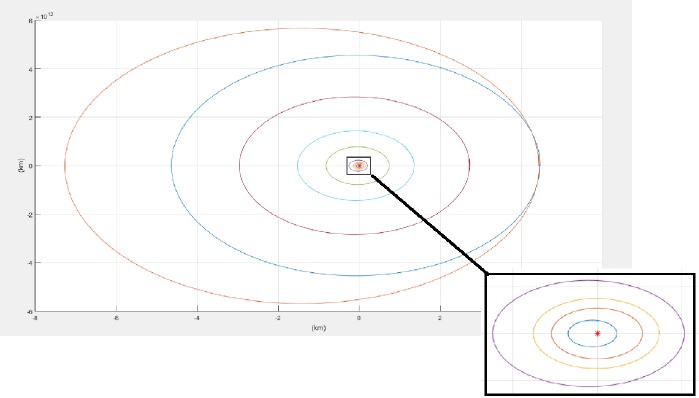}
\caption{\label{fig2}  Closed bound planets' trajectories with and without Yukawa corrections with strength $\a$ nominal.}
\end{figure}

Here, we just solve numerically, using the fourth-order Runge-Kutta method, the Newton's law equation of motion in the c.o.m frame with initial conditions taken from NASA. Thus we solve the equations:
\bea
\ddot{\vec{r}}_1 = G m_p \frac{\vec{r}_2 - \vec{r}_1}{r^3} &,\mbox{Newton},& \ddot{\vec{r}}_2 = G M_{\odot} \frac{\vec{r}_1 - \vec{r}_2}{r^3}= - \frac{M_{\odot}}{m_p}\ddot{\vec{r}}_1 , \nn\\ && \label{num_N} \\
\ddot{\vec{r}}_1 = G m_p \left[(1+\a e^{-\frac{r}{\lambda}}) \frac{1}{r} + \frac{\a}{\lambda}  e^{-\frac{r}{\lambda}} \right] \frac{\vec{r}_2 - \vec{r}_1}{r^2} &,\mbox{Newton+Yukawa},& \ddot{\vec{r}}_2 = - \frac{M_{\odot}}{m_p}\ddot{\vec{r}}_1, \label{num_NYk}
\eea
under the initial conditions given by NASA data of ($r_{min}, v_{max}$):
\bea
\vec{r}_1(t=t_{min}) = \frac{m_p}{m_p+M_{\odot}} \vec{r}_{min} &,& \vec{v}_1(t=t_{min}) = \frac{m_p}{m_p+M_{\odot}} \vec{v}_{max}, \\
\vec{r}_2(t=t_{min}) = -\frac{M_{\odot}}{m_p+M_{\odot}} \vec{r}_{min} &,& \vec{v}_2(t=t_{min}) = -\frac{M_{\odot}}{m_p+M_{\odot}} \vec{v}_{max}.
\eea
Once the trajectory is solved numerically, we check that it is closed, as the Fig. (\ref{fig2}) shows for both the pure Newton and that with the Yukawa corrections (since the differences are not visible on the figure scale containing all the planets). For each obtained orbit, we recalculate the corresponding characteristics ($r_{min}, r_{max}, a, b, e$).

\subsection{\bf Results}

\begin{figure}[tbp]
\hspace{-2.5cm}  
\includegraphics[width=20cm, height=25cm]{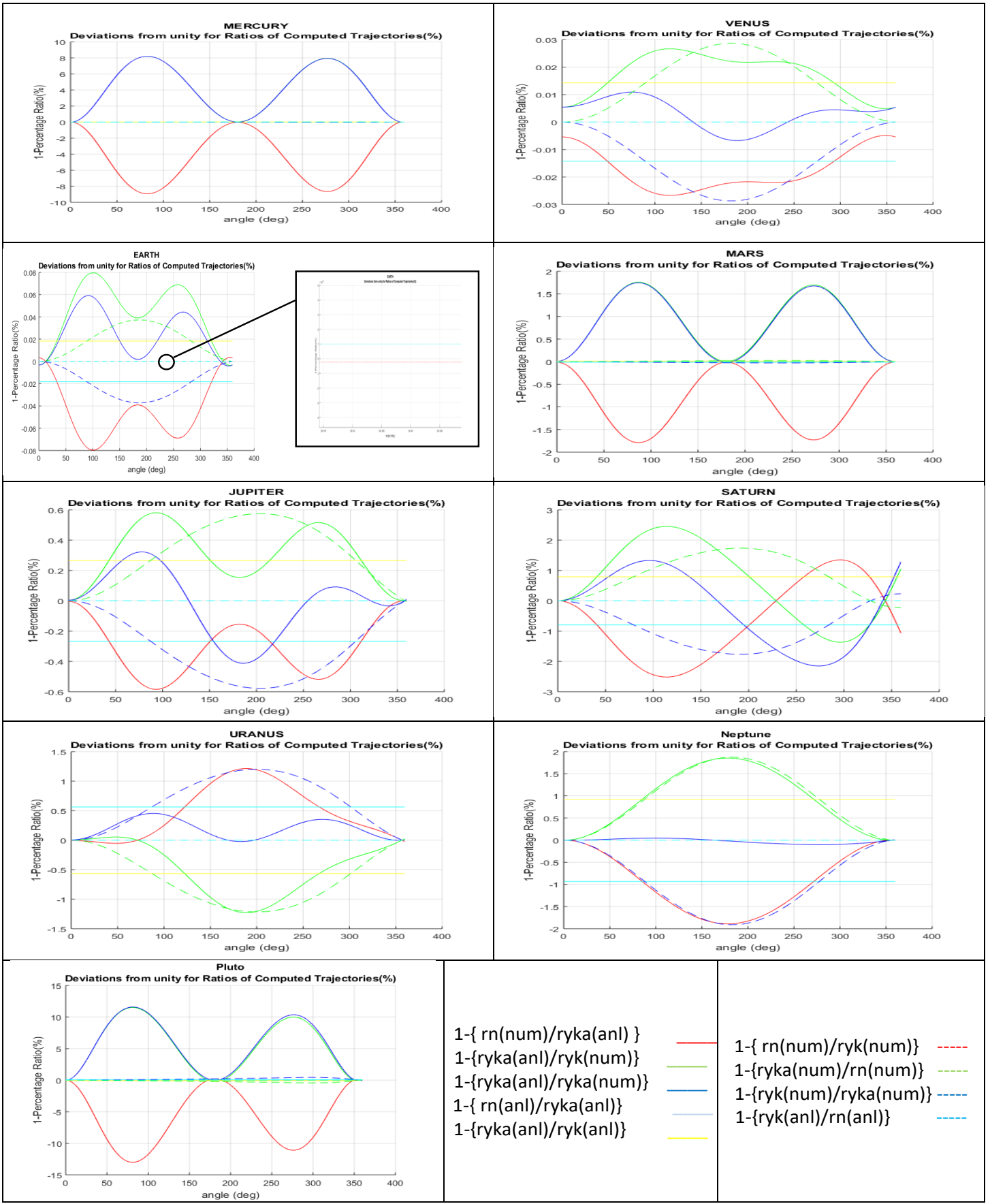}
\vspace{-1.5cm}\caption{\label{fig3}  Deviations from Unity for Ratios of Computed trajectories at each polar angle, according to the considered potential (Newton vs Newton+Yuakawa) and/or to the used method (numerical vs analytical) and/or to the Yukawa strength determination (nominal vs estimated). We show in a zoomed region, for one planet (Earth) generic case, that the dashed red and sky blue curves are very near each other (the same applies to the green and blue curves in Mercury case). }
\end{figure}

We report in the Tables of Appendix A (from \ref{TabAMer-N}  to \ref{TabAPlu-E}),  the calculated characteristics of the resulting trajectories for all the planets in the solar system, corresponding to the pure Newton and the Newton corrected with Yukawa potentials, both in the analytical and the numerical approaches. The odd (even) numbered tables correspond to the nominal (estimated) Yukawa strength $\a$.  The number of moons of each planet is determined according to \cite{12}. Below we explain the meanings of the symbols used in the tables.
\begin{itemize}
\item
$N_{num}$: Numerical calculations using the Newtonian potential.
\item
$N_{anal}$: Analytical calculations using the Newtonian potential.
\item
$R_N=\frac{N_{num}}{N_{anal}}$\%: The percentage ratio of the numerical to the analytical results for Newton potential.
\item
 $(N+YK)_{num}$ : Numerical calculations using the modified potential.
\item
$(N+YK)_{anal}$: Analytical calculations using the modified potential.
\item
$R_{N+YK}=\frac{(N+YK)_{num}}{(N+YK)_{anal}}$\%: The percentage ratio of the numerical to the analytical results for modified potential.
\item
 $R^{N-Obs}_{num}= N_{num}/Obs$\%: Percentage ratio of the numerical results, using the Newtonian potential, to the observed results.
\item
$R^{N-Obs}_{anal}= N_{anal}/Obs$ \%: Percentage ratio the analytical results, using the Newtonian potential, to the observed results.
\item
 $R^{YK-Obs}_{num}=  (N+YK)_{num}/Obs$  \%:  Percentage ratio of the numerical results, using the modified potential, to the observed results.
\item
   $R^{YK-Obs}_{anal}= (N+YK)_{anal}/Obs$ \%: Percentage ratio of the analytical results, using the modified potential, to the observed results.
\end{itemize}
In order to summarize the findings of the Tables, we present in Fig. (\ref{fig3}) plots showing, for each planet and at every polar angle, the deviation from unity of the ratio between two quantities of the following, allowing thus to compare the effects of the considered potential (Newton vs Newton+Yuakawa) and/or the used method (numerical vs analytical) and/or the Yukawa strength determination (nominal vs estimated):
\begin{itemize}
%\item $r(clc)$ representing the geometrical elliptical equation,
\item $rn(num)$ representing the trajectory equation of the numerical approach with Newton potential,
\item $rn(anl)$ representing the trajectory equation of the analytical approach with Newton potential,
\item $ryk(num)$ representing the trajectory equation of the numerical approach with Newton+Yukawa potential and nominal $\a$,
\item $ryk(anl)$ representing the trajectory equation of the analytical approach with Newton+Yukawa potential and nominal $\a$,
\item $ryka(num)$ representing the trajectory equation of the numerical approach with Newton+Yukawa potential and estimated $\a$,
\item $ryka(anl)$ representing the trajectory equation of the analytical approach with Newton+Yukawa potential and estimated $\a$.
\end{itemize}

We see that some ratios (e.g. the dashed red and sky blue) do coincide near zero deviation from one, meaning no tangible effect of adding the Yukawa correction, be it in the analytic or the numeric method, as long as one takes the nominal value of $\alpha$. Also,we note local extremums for the deviations from unity at polar angles multiples of $\pi/2$ as a generic feature in many plots. One can interpret the large values of the deviations for the nearest (Mercury) and the farthest (Pluto) planet, in that for the former; the perturbative effect of solar winds, important as we approach the sun, was not taken into consideration, whereas for the furthest; accumulating gravitational screening effects of the other planets and their moons, which were not considered in the study, are becoming important especially for a small sized- planetoid like Pluto.

\begin{figure}[tbp]
\hspace{-2.5cm}
\includegraphics[width=20cm, height=25cm]{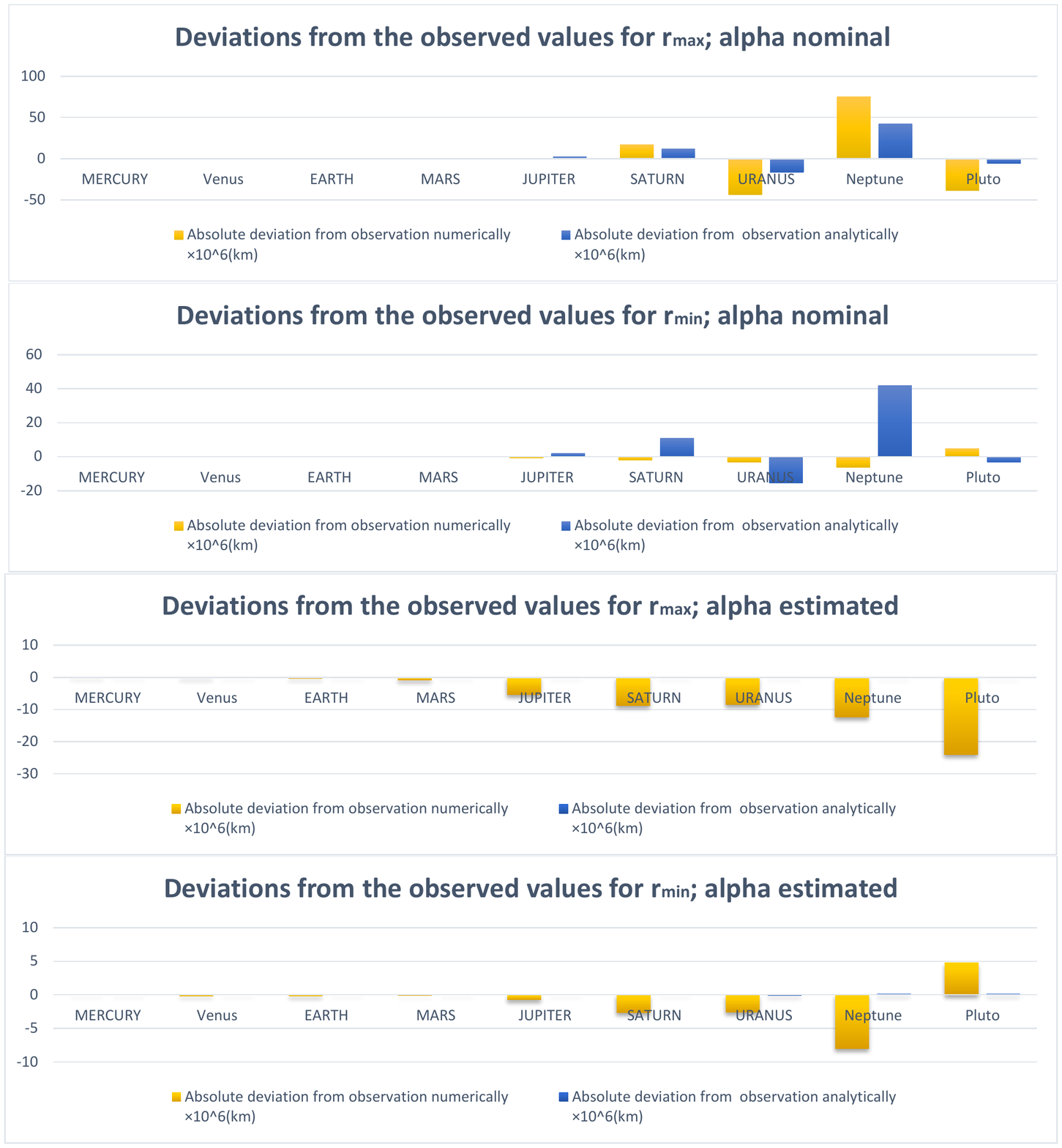}
\vspace{-2.5cm}
\caption{\label{fig4}  Absolute deviations from observed data for each planet, according to the used method (numerical vs analytical) and/or to the Yukawa strength determination (nominal vs estimated).}
\end{figure}

In order to show the effects of the separating distance effect, one should compute the absolute deviations from observed data for each planet. In appendix B, the Tables \ref{TabBrmax-N}, \ref{TabBrmin-N} (\ref{TabBrmax-E}, \ref{TabBrmin-E}), report the deviation from observation for each planet of $r_{max}, r_{min}$ respectively, in the case of nominal (estimated) $\a$. We summarize these findings in Fig. (\ref{fig4}).
We see that the agreement between the numerical and analytical solutions is excellent in both estimated and nominal $\a$ cases. We see that the deviations due to Yukawa correction are not large, but note the following:
\begin{enumerate}
\item {\underline{For estimated $\a$}:}
\begin{itemize}
\item $r_{max}$-deviation: The numerical deviation is larger by about $10^{3}$ times the analytical deviation. In general, it increases the further the planet is, and reaches a maximum of order ($-25$ million km) (less than the observed value) in Pluto.
\item $r_{min}$-deviation: Again, the numerical deviation is larger by about ($10^1$-$10^2$)-order of magnitude than analytical deviation, where it is largest in Neptune ($-8$ million km), however it reverses sign and becomes ($+5$ million km) more than the observation in Pluto.
\end{itemize} 
\item {\underline{For nominal $\a$}:} 
\begin{itemize}
\item $r_{max}$-deviation: The numerical deviation is larger than the analytical one, but are of the same order reaching a maximum of $+80$ ($+40$) million km using the numerical (analytical) method in Neptune. For Pluto and Uranus, we get ($-40$) million km in the numerical method (less than observed).
\item $r_{min}$-deviation: The analytical deviation is larger, and sometimes reverses sign compared to the numerical. For example, in Neptune the analytical approach gives a deviation of (+40 million km) from observation, whereas the numerical one gives a deviation of $-5$ million km (less than the observed value).
\end{itemize}
\end{enumerate}  
 Actually, the disagreements with observations are due to several reasons: the first one is physical in nature, in that it results from neglecting the perturbation due to third bodies, or, more generally, the effect of the natural satellites, such as moons or asteroids. Also, we did not either take into account the radiation and the solar wind physical effects. Moreover, the results were obtained as a 2-body problem, and hence the movement of more distant planets might be affected by planets closer to the sun, which can be present not in the dominant term, but in higher orders of expansion. The second factor lies in the computational side, and concerns the numerical method used, the value of the step size, and the high sensitivity of the problem to the initial conditions. One should also mention that for the analytical solution we restricted the study to leading order neglecting higher orders in the expansion of exponentials, whereas for the numerical solution the entire exponential is considered.

%%%%%%%%%%%%%%%%%%%%%%%%%%%%%%%%%%%%%%%%%%
\section{Summary  and Conclusion}
In this work, we followed \cite{5} and used the Hamilton's formulation in order to obtain the differential equation of motion and the path equation for the gravitational two-body system. The developments are carried out in the case of the pure Newtonian potential, the Newtonian corrected with Yukawa type potential and the pure Yukawa potential. As in \cite{5}, we have reviewed the stability problem, constructed the linearization matrix and tested the stability of the system for a Yukawa correction, and found that it is of a central solution type, which implies stable solutions near the fixed point. We repeated the analysis for a purely Yukawa force and found similar results. We also confirmed that the modified potential obeys the Bertrand’s theorem.

Then, we determined the parameters' set corresponding to the planets of the solar system starting from the observed ($r_{min}, v_{max}, e$) estimating $\a$. For both the estimated and nominal values of $\a$, we determined the characteristics of the trajectories numerically and analytically, and compared between the methods and with the observed data. We explained the extent to which these results are consistent with the observational data, presenting in form of histograms the absolute deviations from observations, which were found to give an upper deviation of order 80 million km in Neptune using nominal $\a$, and 20 million km in Pluto using estimated $\a$.
\vspace{1cm}
\begin{comment}
\authorcontributions{``All authors contributed equally to this work.''}

\funding{``This research received no external funding.''}

\institutionalreview{}

\informedconsent{``Informed consent was obtained from all subjects involved in the study.''}

\dataavailability{}
\end{comment}

{\bf Acknowledgments: }{N. Chamoun acknowledges support from the ICTP-Associate program, from the Humboldt Foundation and from the CAS-PIFI scholarship.}

%\conflictsofinterest{``The authors declare no conflict of interest.''}

\begin{appendices}

\appendix{\bf A. Tables of Calculated/Observed Parameters of the Planets}
\clearpage
\setcounter{table}{0}
\renewcommand{\thetable}{A\arabic{table}}

\begin{table}[]
\begin{tabular}{|p{40mm}|l|l|l|l|l|}
\hline
Mercury  &   $r_{min}$($\times 10^6 km$)          &   $r_{max}$($\times 10^6 km$)          &   $a$($\times 10^6 km$)          &    $b$($\times 10^6 km$)         &      eccentricity       \\ \hline \hline
 $N_{num}$       & 46          & 69.832 & 57.916 & 56.67679158 & 0.205744228 \\ \hline
  $N_{anal}$      & 47 & 72.043 & 59.756 & 58.47840586 & 0.205646344 \\ \hline
 $R_N=\frac{N_{num}}{N_{anal}}$\%        & 97  & 96.930 & 96.921 & 96.91918025 & 99.95242442 \\ \hline
  $(N+YK)_{num}$      & 46          & 69.831 & 57.916 & 56.67678243 & 0.205738221 \\ \hline
  $(N+YK)_{anal}$      & 47 & 72.043 & 59.756 & 58.47840528 & 0.205646344 \\ \hline
 $R_{N+YK}$\\$=\frac{(N+YK)_{num}}{(N+YK)_{anal}}$\%        & 97 & 96.930 & 96.921 & 96.91916556 & 99.95534277 \\ \hline
    Observation    & 46          & 69.818      & 57.909      & \_\_\_\_\_  & 0.20563069  \\ \hline
 $R^{N-Obs}_{num}$\\$= N_{num}/Obs$ \%    & 100         & 99.980 & 99.988 & \_\_\_\_\_  & 99.94481595 \\ \hline
    $R^{N-Obs}_{anal}$\\$= N_{anal}/Obs$ \%      & 97  & 96.911 & 96.909 & \_\_\_\_\_  & 99.9923879  \\ \hline
  $R^{YK-Obs}_{num}=$\\$  (N+YK)_{num}/Obs$  \%   & 100         & 99.981 & 99.988 & \_\_\_\_\_  & 99.94773407 \\ \hline
   $R^{YK-Obs}_{anal}=$\\$ (N+YK)_{anal}/Obs$  \%         & 97 & 96.911  & 96.909 & \_\_\_\_\_  & 99.9923879  \\ \hline
   \multicolumn{6}{|c|}{nominal $\a=10^{-8}$}\\ \hline
\end{tabular}
\caption{ The values of the calculated and observational astronomical parameters of the planet Mercury whose number of moons is 0} \label{TabAMer-N}
\end{table}

%%%%%%%%%%%%%%%%%%%%%%%%%%%%%%
\begin{table}[]
\begin{tabular}{|p{40mm}|l|l|l|l|l|}
\hline
Mercury  &   $r_{min}$($\times 10^6 km$)          &   $r_{max}$($\times 10^6 km$)          &   $a$($\times 10^6 km$)          &    $b$($\times 10^6 km$)         &      eccentricity       \\ \hline \hline
 $N_{num}$       & 46          & 69.623 & 57.826 & 56.65144795 & 0.2022 \\ \hline
  $N_{anal}$      & 46 & 69.819 & 57.912 & 56.67470066 & 0.2056 \\ \hline
 $R_N=\frac{N_{num}}{N_{anal}}$\%        & 100  & 99.719 & 99.851 & 99.95897162 & 98.3743 \\ \hline
  $(N+YK)_{num}$      & 46          & 69.613 & 57.820 & 56.64729064 & 0.2022 \\ \hline
  $(N+YK)_{anal}$      & 46 & 69.815 & 57.908 & 56.6714469 & 0.2056 \\ \hline
 $R_{N+YK}$\\$=\frac{(N+YK)_{num}}{(N+YK)_{anal}}$\%        & 100 &99.711 & 99.847 & 99.9573749 & 98.3408 \\ \hline
    Observation    & 46          & 69.818      & 57.909      & \_\_\_\_\_  & 0.2056  \\ \hline
 $R^{N-Obs}_{num}$\\$= N_{num}/Obs$ \%    & 100         & 99.721 & 99.856 & \_\_\_\_\_  & 98.3817 \\ \hline
    $R^{N-Obs}_{anal}$\\$= N_{anal}/Obs$ \%      & 100  & 100.001 & 100.005 & \_\_\_\_\_  & 100.0075  \\ \hline
  $R^{YK-Obs}_{num}=$\\$  (N+YK)_{num}/Obs$  \%   & 100         & 99.706 & 99.847 & \_\_\_\_\_  & 98.3482 \\ \hline
   $R^{YK-Obs}_{anal}=$\\$ (N+YK)_{anal}/Obs$  \%         & 100 & 99.995  & 99.999 & \_\_\_\_\_  & 100.0075  \\ \hline
    \multicolumn{6}{|c|}{estimated $\a=5.741444131301954\times10^{-5}$}\\ \hline
    \end{tabular}
\caption{ The values of the calculated and observational astronomical parameters of the planet Mercury whose number of moons is 0} \label{TabAMer-E}
\end{table}
%%%%%%%%%%%%%%%%%%%%%%%%%%%%%%
%%%%%%%%%%%%%%%%%%%%%%%%%%%%%%%

\begin{table}[]
\begin{tabular}{|p{40mm}|l|l|l|l|l|}
\hline
Venus  &   $r_{min}$($\times 10^6 km$)          &   $r_{max}$($\times 10^6 km$)          &   $a$($\times 10^6 km$)          &    $b$($\times 10^6 km$)         &      eccentricity       \\ \hline \hline
 $N_{num}$       & 107.30          & 108.689 & 107.99 & 107.9982364 & 0.0044 \\ \hline
  $N_{anal}$      & 107.48 & 108.961 & 108.22 & 108.2222348 & 0.0072 \\ \hline
 $R_N=\frac{N_{num}}{N_{anal}}$\%        & 99.83  & 99.750 & 99.79 & 99.79301998 & 60.8187 \\ \hline
  $(N+YK)_{num}$      & 107.30          & 108.689 & 107.99 & 107.9982353 & 0.0044 \\ \hline
  $(N+YK)_{anal}$      & 107.48 & 108.961 & 108.22 & 108.2222337 & 0.0072 \\ \hline
 $R_{N+YK}$\\$=\frac{(N+YK)_{num}}{(N+YK)_{anal}}$\%        & 99.83 & 99.750 & 99.79 & 99.79301998 & 60.8189 \\ \hline
    Observation    & 107.48          & 108.941     & 108.21      & \_\_\_\_\_  &0.0068  \\ \hline
 $R^{N-Obs}_{num}$\\$= N_{num}/Obs$ \%    & 99.83         & 99.769 & 99.80 & \_\_\_\_\_  & 64.7150 \\ \hline
    $R^{N-Obs}_{anal}$\\$= N_{anal}/Obs$ \%      & 100.00  & 100.018 & 100.01 & \_\_\_\_\_  & 106.4063  \\ \hline
  $R^{YK-Obs}_{num}=$\\$  (N+YK)_{num}/Obs$  \%   & 99.83         & 99.769 & 99.80 & \_\_\_\_\_  & 64.7152 \\ \hline
   $R^{YK-Obs}_{anal}=$\\$ (N+YK)_{anal}/Obs$  \%         & 100.00 & 100.018  & 100.01 & \_\_\_\_\_  & 106.4063 \\ \hline
   \multicolumn{6}{|c|}{nominal $\a=10^{-8}$}\\ \hline
\end{tabular}
\caption{ The values of the calculated and observational astronomical parameters of the planet Venus whose number of moons is 0} \label{TabAVen-N}
\end{table}
%%%%%%%%%%%%%%%%%%%%%%%%%%%%%%%%%
\begin{table}[]
\begin{tabular}{|p{40mm}|l|l|l|l|l|}
\hline
Venus  &   $r_{min}$($\times 10^6 km$)          &   $r_{max}$($\times 10^6 km$)          &   $a$($\times 10^6 km$)          &    $b$($\times 10^6 km$)         &      eccentricity       \\ \hline \hline
 $N_{num}$       & 107.30          & 108.689 & 107.99 & 107.9982364 & 0.0044 \\ \hline
  $N_{anal}$      & 107.48 & 108.961 & 108.22 & 108.2222348 & 0.0072 \\ \hline
 $R_N=\frac{N_{num}}{N_{anal}}$\%        & 99.83  & 99.750 & 99.79 & 99.79301998 & 60.8187 \\ \hline
  $(N+YK)_{num}$      & 107.30          & 108.658 & 107.98 & 107.9821956 & 0.0045 \\ \hline
  $(N+YK)_{anal}$      & 107.47 & 108.945 & 108.20 & 108.2068155 & 0.0072 \\ \hline
 $R_{N+YK}$\\$=\frac{(N+YK)_{num}}{(N+YK)_{anal}}$\%        & 99.84 & 99.736 & 99.78 & 99.79241613 & 63.0300 \\ \hline
    Observation    & 107.48          & 108.941     & 108.21      & \_\_\_\_\_  &0.0068  \\ \hline
 $R^{N-Obs}_{num}$\\$= N_{num}/Obs$ \%    & 99.83         & 99.769 & 99.80 & \_\_\_\_\_  & 64.7150 \\ \hline
    $R^{N-Obs}_{anal}$\\$= N_{anal}/Obs$ \%      & 100.00  & 100.018 & 100.01 & \_\_\_\_\_  & 106.4063  \\ \hline
  $R^{YK-Obs}_{num}=$\\$  (N+YK)_{num}/Obs$  \%   & 99.83        &99.740 & 99.78 & \_\_\_\_\_  & 67.0680 \\ \hline
   $R^{YK-Obs}_{anal}=$\\$ (N+YK)_{anal}/Obs$  \%         & 99.99 & 100.004  & 99.99 & \_\_\_\_\_  & 106.4063 \\ \hline
   \multicolumn{6}{|c|}{estimated $\a=1.424988220126711\times10^{-4}$}\\ \hline
   \end{tabular}
\caption{ The values of the calculated and observational astronomical parameters of the planet Venus whose number of moons is 0} \label{TabAVen-E}
\end{table}
%%%%%%%%%%%%%%%%%%%%%%%%%%%%
%%%%%%%%%%%%%%%%%%%%%%%%%%%%
\begin{table}[]
\begin{tabular}{|p{40mm}|l|l|l|l|l|}
\hline
EARTH  &   $r_{min}$($\times 10^6 km$)          &   $r_{max}$($\times 10^6 km$)          &   $a$($\times 10^6 km$)          &    $b$($\times 10^6 km$)         &      eccentricity       \\ \hline \hline
 $N_{num}$       & 146.884          & 151.7 & 149.336 & 149.319847 & 0.0156 \\ \hline
  $N_{anal}$      & 147.126 & 152.1 & 149.625 & 149.6034965 &  0.0168 \\ \hline
 $R_N=\frac{N_{num}}{N_{anal}}$\%        & 99.835  & 99.7  & 99.806 & 99.81039915 & 92.5721 \\ \hline
  $(N+YK)_{num}$      & 146.884          & 151.7 & 149.336 & 149.3198455 & 0.0156 \\ \hline
  $(N+YK)_{anal}$      &147.126 & 152.1 & 149.625 & 149.603495 & 0.0168 \\ \hline
 $R_{N+YK}$\\$=\frac{(N+YK)_{num}}{(N+YK)_{anal}}$\%        & 99.835  & 99.7 & 99.806 & 99.81039915 &  92.5720 \\ \hline
    Observation    & 147.095          & 152.1    & 149.598      & \_\_\_\_\_  &  0.0167 \\ \hline
 $R^{N-Obs}_{num}$\\$= N_{num}/Obs$ \%    & 99.8572         & 99.7 & 99.825 & \_\_\_\_\_  & 93.5903 \\ \hline
    $R^{N-Obs}_{anal}$\\$= N_{anal}/Obs$ \%      & 99.978  & 99.9 & 99.981 & \_\_\_\_\_  & 98.9120  \\ \hline
  $R^{YK-Obs}_{num}=$\\$  (N+YK)_{num}/Obs$  \%   & 99.857         & 99.7 & 99.825 & \_\_\_\_\_  & 93.5902   \\ \hline
   $R^{YK-Obs}_{anal}=$\\$ (N+YK)_{anal}/Obs$  \%         & 99.978  & 99.9  & 99.981 & \_\_\_\_\_  & 98.9120 \\ \hline
   \multicolumn{6}{|c|}{nominal $\a=10^{-8}$}\\ \hline
\end{tabular}
\caption{ The values of the calculated and observational astronomical parameters of the planet Earth whose number of moons is 0} \label{TabAEar-N}
\end{table}
%%%%%%%%%%%%%%%%%%%%%%%%%%
\begin{table}[]
\begin{tabular}{|p{40mm}|l|l|l|l|l|}
\hline
EARTH  &   $r_{min}$($\times 10^6 km$)          &   $r_{max}$($\times 10^6 km$)          &   $a$($\times 10^6 km$)          &    $b$($\times 10^6 km$)         &      eccentricity       \\ \hline \hline
 $N_{num}$       & 146.884          & 151.7 & 149.336 & 149.319847 & 0.0156 \\ \hline
  $N_{anal}$      & 147.126 & 152.1 & 149.625 & 149.6034965 &  0.0168 \\ \hline
 $R_N=\frac{N_{num}}{N_{anal}}$\%        & 99.835  & 99.7  & 99.806 & 99.81039915 & 92.5721 \\ \hline
  $(N+YK)_{num}$      & 146.883          & 151.7 & 149.307 & 149.2910008 & 0.0154 \\ \hline
  $(N+YK)_{anal}$      &147.099 & 152.0 & 149.597 & 149.5762082 & 0.01688 \\ \hline
 $R_{N+YK}$\\$=\frac{(N+YK)_{num}}{(N+YK)_{anal}}$\%        & 99.853  & 99.7 & 99.805 & 99.80932302 & 91.4700 \\ \hline
    Observation    & 147.095          & 152.1    & 149.598      & \_\_\_\_\_  &  0.0167 \\ \hline
 $R^{N-Obs}_{num}$\\$= N_{num}/Obs$ \%    & 99.8572         & 99.7 & 99.825 & \_\_\_\_\_  & 93.5903 \\ \hline
    $R^{N-Obs}_{anal}$\\$= N_{anal}/Obs$ \%      & 99.978  & 99.9 & 99.981 & \_\_\_\_\_  & 98.9120  \\ \hline
  $R^{YK-Obs}_{num}=$\\$  (N+YK)_{num}/Obs$  \%   & 99.856         & 99.7 & 99.805 & \_\_\_\_\_  & 92.4761   \\ \hline
   $R^{YK-Obs}_{anal}=$\\$ (N+YK)_{anal}/Obs$  \%         & 100.003  & 99.9  & 99.999 & \_\_\_\_\_  & 101.0999 \\ \hline
   \multicolumn{6}{|c|}{estimated $\a=1.824376359731428\times10^{-4}$}\\ \hline
   \end{tabular}
\caption{ The values of the calculated and observational astronomical parameters of the planet Earth whose number of moons is 0} \label{TabAEar-E}
\end{table}
%%%%%%%%%%%%%%%%%%%%%%%
%%%%%%%%%%%%%%%%%%%%%%
\begin{table}[]
\begin{tabular}{|p{40mm}|l|l|l|l|l|}
\hline
MARS  &   $r_{min}$($\times 10^6 km$)          &   $r_{max}$($\times 10^6 km$)          &   $a$($\times 10^6 km$)          &    $b$($\times 10^6 km$)         &      eccentricity       \\ \hline \hline
 $N_{num}$       & 206.57          & 248.480 & 227.52 & 226.6509159 & 0.0898 \\ \hline
  $N_{anal}$      & 206.64 & 249.277 & 227.96 & 226.9631182 & 0.0935 \\ \hline
 $R_N=\frac{N_{num}}{N_{anal}}$\%        & 99.96  & 99.680 & 99.80 & 99.8624436 & 96.0965 \\ \hline
  $(N+YK)_{num}$      & 206.57          & 248.480 & 227.52 & 226.6509134 & 0.0898 \\ \hline
  $(N+YK)_{anal}$      & 206.64 & 249.277 & 227.96 &  226.9631159 & 0.0935 \\ \hline
 $R_{N+YK}$\\$=\frac{(N+YK)_{num}}{(N+YK)_{anal}}$\%        & 99.96 & 99.680 & 99.80 & 99.86244351 & 96.0965 \\ \hline
    Observation    & 206.65          & 249.261    & 227.94      & \_\_\_\_\_  &  0.0935 \\ \hline
 $R^{N-Obs}_{num}$\\$= N_{num}/Obs$ \%    & 99.96        & 99.687 & 99.81 & \_\_\_\_\_  & 96.1252 \\ \hline
    $R^{N-Obs}_{anal}$\\$= N_{anal}/Obs$ \%      & 99.99  & 100.006 & 100.01 & \_\_\_\_\_  & 100.0298  \\ \hline
  $R^{YK-Obs}_{num}=$\\$  (N+YK)_{num}/Obs$  \%   & 99.96        & 99.687 & 99.81 & \_\_\_\_\_  &   96.1252 \\ \hline
   $R^{YK-Obs}_{anal}=$\\$ (N+YK)_{anal}/Obs$  \%         & 99.99 & 100.006  & 100.01 & \_\_\_\_\_  & 100.0298 \\ \hline
   \multicolumn{6}{|c|}{nominal $\a=10^{-8}$}\\ \hline
\end{tabular}
\caption{ The values of the calculated and observational astronomical parameters of the planet Mars whose number of moons is 0} \label{TabAMar-N}
\end{table}
%%%%%%%%%%%%%%%%%%%%%%%%
\begin{table}[]
\begin{tabular}{|p{40mm}|l|l|l|l|l|}
\hline
MARS  &   $r_{min}$($\times 10^6 km$)          &   $r_{max}$($\times 10^6 km$)          &   $a$($\times 10^6 km$)          &    $b$($\times 10^6 km$)         &      eccentricity       \\ \hline \hline
 $N_{num}$       & 206.57          & 248.480 & 227.52 & 226.6509159 & 0.0898 \\ \hline
  $N_{anal}$      & 206.64 & 249.277 & 227.96 & 226.9631182 & 0.0935 \\ \hline
 $R_N=\frac{N_{num}}{N_{anal}}$\%        & 99.96  & 99.680 & 99.80 & 99.8624436 & 96.0965 \\ \hline
  $(N+YK)_{num}$      & 206.57          & 248.425 & 227.49 & 226.6249874 & 0.0897 \\ \hline
  $(N+YK)_{anal}$      & 206.62 & 249.252 & 227.93 &  226.9402451 & 0.0935 \\ \hline
 $R_{N+YK}$\\$=\frac{(N+YK)_{num}}{(N+YK)_{anal}}$\%        & 99.97 & 99.668 & 99.80 & 99.86108339 & 95.9861 \\ \hline
    Observation    & 206.65          & 249.261    & 227.94      & \_\_\_\_\_  &  0.0935 \\ \hline
 $R^{N-Obs}_{num}$\\$= N_{num}/Obs$ \%    & 99.96        & 99.687 & 99.81 & \_\_\_\_\_  & 96.1252 \\ \hline
    $R^{N-Obs}_{anal}$\\$= N_{anal}/Obs$ \%      & 99.99  & 100.006 & 100.01 & \_\_\_\_\_  & 100.0298  \\ \hline
  $R^{YK-Obs}_{num}=$\\$  (N+YK)_{num}/Obs$  \%   & 99.96        & 99.664 & 99.80 & \_\_\_\_\_  &   96.0147 \\ \hline
   $R^{YK-Obs}_{anal}=$\\$ (N+YK)_{anal}/Obs$  \%         & 99.98 & 99.996 & 99.99 & \_\_\_\_\_  & 100.0298 \\ \hline
   \multicolumn{6}{|c|}{estimated $\a=1.007889331583467\times10^{-4}$}\\ \hline
   \end{tabular}
\caption{ The values of the calculated and observational astronomical parameters of the planet Mars whose number of moons is 0} \label{TabAMar-E}
\end{table}
%%%%%%%%%%%%%%%%%%%%
%%%%%%%%%%%%%%%%%%%%%%
\begin{table}[]
\begin{tabular}{|p{40mm}|l|l|l|l|l|}
\hline
JUPITER  &   $r_{min}$($\times 10^6 km$)          &   $r_{max}$($\times 10^6 km$)          &   $a$($\times 10^6 km$)          &    $b$($\times 10^6 km$)         &      eccentricity       \\ \hline \hline
 $N_{num}$       & 739.902          & 815.533 & 777.717 & 776.9190412 & 0.0469 \\ \hline
  $N_{anal}$      & 742.542 & 818.568 & 780.555 & 779.626266 & 0.04873 \\ \hline
 $R_N=\frac{N_{num}}{N_{anal}}$\%        & 99.644 & 99.629 & 99.636 & 99.65275352 & 96.3711 \\ \hline
  $(N+YK)_{num}$      & 739.902          & 815.533 & 777.717 & 776.9190329 & 0.0469 \\ \hline
  $(N+YK)_{anal}$      & 742.542 & 818.568 & 780.555 &  779.6262582 & 0.0487 \\ \hline
 $R_{N+YK}$\\$=\frac{(N+YK)_{num}}{(N+YK)_{anal}}$\%        & 99.644 & 99.629 & 99.636 & 99.65275345 & 96.3711 \\ \hline
    Observation    & 740.595          & 816.363    & 778.479      & \_\_\_\_\_  &  0.0487 \\ \hline
 $R^{N-Obs}_{num}$\\$= N_{num}/Obs$ \%    & 99.906        & 99.898 & 99.902 & \_\_\_\_\_  & 96.4399  \\ \hline
    $R^{N-Obs}_{anal}$\\$= N_{anal}/Obs$ \%      & 100.262  & 100.270 & 100.266 & \_\_\_\_\_  & 100.0714  \\ \hline
  $R^{YK-Obs}_{num}=$\\$  (N+YK)_{num}/Obs$  \%   & 99.906        & 99.898 & 99.902 & \_\_\_\_\_  &   96.4399 \\ \hline
   $R^{YK-Obs}_{anal}=$\\$ (N+YK)_{anal}/Obs$  \%         & 100.262 & 100.270 & 100.266 & \_\_\_\_\_  & 100.0714\\ \hline
   \multicolumn{6}{|c|}{nominal $\a=10^{-8}$}\\ \hline
\end{tabular}
\caption{ The values of the calculated and observational astronomical parameters of the planet Jupiter whose number of moons is 0} \label{TabAJup-N}
\end{table}
%%%%%%%%%%%%%%%%%%%
\begin{table}[]
\begin{tabular}{|p{40mm}|l|l|l|l|l|}
\hline
JUPITER  &   $r_{min}$($\times 10^6 km$)          &   $r_{max}$($\times 10^6 km$)          &   $a$($\times 10^6 km$)          &    $b$($\times 10^6 km$)         &      eccentricity       \\ \hline \hline
 $N_{num}$       & 739.902          & 815.533 & 777.717 & 776.9190412 & 0.0469 \\ \hline
  $N_{anal}$      & 742.542 & 818.568 & 780.555 & 779.626266 & 0.04873 \\ \hline
 $R_N=\frac{N_{num}}{N_{anal}}$\%        & 99.644 & 99.629 & 99.636 & 99.65275352 & 96.3711 \\ \hline
  $(N+YK)_{num}$      & 739.837          & 810.932 & 775.385 & 774.6852056 & 0.0441 \\ \hline
  $(N+YK)_{anal}$      & 740.567 & 816.390 & 778.478 &  777.5526264 & 0.0487 \\ \hline
 $R_{N+YK}$\\$=\frac{(N+YK)_{num}}{(N+YK)_{anal}}$\%        & 99.901 & 99.331 & 99.602 & 99.63122486 & 90.6263 \\ \hline
    Observation    & 740.595          & 816.363    & 778.479      & \_\_\_\_\_  &  0.0487 \\ \hline
 $R^{N-Obs}_{num}$\\$= N_{num}/Obs$ \%    & 99.906        & 99.898 & 99.902 & \_\_\_\_\_  & 96.4399  \\ \hline
    $R^{N-Obs}_{anal}$\\$= N_{anal}/Obs$ \%      & 100.262  & 100.270 & 100.266 & \_\_\_\_\_  & 100.0714  \\ \hline
  $R^{YK-Obs}_{num}=$\\$  (N+YK)_{num}/Obs$  \%   & 99.897        & 99.334 & 99.602 & \_\_\_\_\_  &   90.6911 \\ \hline
   $R^{YK-Obs}_{anal}=$\\$ (N+YK)_{anal}/Obs$  \%         & 99.996 & 100.003  & 99.999 & \_\_\_\_\_  & 100.0714\\ \hline
    \multicolumn{6}{|c|}{estimated $\a=2.666880127522\times10^{-3}$}\\ \hline
    \end{tabular}
\caption{ The values of the calculated and observational astronomical parameters of the planet Jupiter whose number of moons is 0} \label{TabAJup-E}
\end{table}
%%%%%%%%%%%%%%%
%%%%%%%%%%%%%%%%%
\begin{table}[]
\begin{tabular}{|p{40mm}|l|l|l|l|l|}
\hline
SATURN  &   $r_{min}$($\times 10^6 km$)          &   $r_{max}$($\times 10^6 km$)          &   $a$($\times 10^6 km$)          &    $b$($\times 10^6 km$)         &      eccentricity       \\ \hline \hline
 $N_{num}$       & 1355.461         & 1523.344 &   1439.403 & 1437.455093 & 0.055   \\ \hline
  $N_{anal}$      & 1368.378 & 1518.496 & 1443.437 & 1441.481829 & 0.052 \\ \hline
 $R_N=\frac{N_{num}}{N_{anal}}$\%        & 99.056  & 100.319 & 99.720 & 99.72065302 & 106.042 \\ \hline
  $(N+YK)_{num}$      & 1355.461          & 1523.344 & 1439.403 & 1437.455078 & 0.055 \\ \hline
  $(N+YK)_{anal}$      & 1368.378 & 1518.496 & 1443.437 &  1441.481815 & 0.052 \\ \hline
 $R_{N+YK}$\\$=\frac{(N+YK)_{num}}{(N+YK)_{anal}}$\%        & 99.056 & 100.319 & 99.720 & 99.72065294 & 106.042 \\ \hline
    Observation    & 1357.554         & 1506.527    & 1432.041      & \_\_\_\_\_  &  0.052 \\ \hline
 $R^{N-Obs}_{num}$\\$= N_{num}/Obs$ \%    & 99.845        & 101.116 & 100.514 & \_\_\_\_\_  & 106.081 \\ \hline
    $R^{N-Obs}_{anal}$\\$= N_{anal}/Obs$ \%      & 100.797  & 100.794 & 100.795 & \_\_\_\_\_  & 100.036  \\ \hline
  $R^{YK-Obs}_{num}=$\\$  (N+YK)_{num}/Obs$  \%   & 99.845        &101.116 &   100.514   & \_\_\_\_\_  &   106.081 \\ \hline
   $R^{YK-Obs}_{anal}=$\\$ (N+YK)_{anal}/Obs$  \%         & 100.797 & 100.794  & 100.795   & \_\_\_\_\_  &   100.036 \\ \hline
   \multicolumn{6}{|c|}{nominal $\a=10^{-8}$}\\ \hline
\end{tabular}
\caption{ The values of the calculated and observational astronomical parameters of the planet Saturn whose number of moons is 0} \label{TabASat-N}
\end{table}
%%%%%%%%%%%%%%%%%%%%
\begin{table}[]
\begin{tabular}{|p{40mm}|l|l|l|l|l|}
\hline
SATURN  &   $r_{min}$($\times 10^6 km$)          &   $r_{max}$($\times 10^6 km$)          &   $a$($\times 10^6 km$)          &    $b$($\times 10^6 km$)         &      eccentricity       \\ \hline \hline
 $N_{num}$       & 1355.461         & 1523.344 &   1439.403 & 1437.455093 & 0.055   \\ \hline
  $N_{anal}$      & 1368.378 & 1518.496 & 1443.437 & 1441.481829 & 0.052 \\ \hline
 $R_N=\frac{N_{num}}{N_{anal}}$\%        & 99.056  & 100.319 & 99.720 & 99.72065302 & 106.042 \\ \hline
  $(N+YK)_{num}$      & 1354.869          & 1497.652 & 1426.261 & 1424.954776 & 0.046 \\ \hline
  $(N+YK)_{anal}$      & 1357.574 & 1506.507 & 1432.040 &  1430.100672 & 0.052 \\ \hline
 $R_{N+YK}$\\$=\frac{(N+YK)_{num}}{(N+YK)_{anal}}$\%        & 99.800 & 99.412 & 99.596 & 99.64017246 & 89.244 \\ \hline
    Observation    & 1357.554         & 1506.527    & 1432.041      & \_\_\_\_\_  &  0.052 \\ \hline
 $R^{N-Obs}_{num}$\\$= N_{num}/Obs$ \%    & 99.845        & 101.116 & 100.514 & \_\_\_\_\_  & 106.081 \\ \hline
    $R^{N-Obs}_{anal}$\\$= N_{anal}/Obs$ \%      & 100.797  & 100.794 & 100.795 & \_\_\_\_\_  & 100.036  \\ \hline
  $R^{YK-Obs}_{num}=$\\$  (N+YK)_{num}/Obs$  \%   & 99.802        & 99.410 &   99.596   & \_\_\_\_\_  &   89.277 \\ \hline
   $R^{YK-Obs}_{anal}=$\\$ (N+YK)_{anal}/Obs$  \%         & 100.001 & 99.998  & 99.999   & \_\_\_\_\_  &   100.036 \\ \hline
   \multicolumn{6}{|c|}{estimated $\a=7.958291053541\times10^{-3}$}\\ \hline
   \end{tabular}
\caption{ The values of the calculated and observational astronomical parameters of the planet Saturn whose number of moons is 0} \label{TabASat-E}
\end{table}
%%%%%%%%%%%%%%%%%%%%%
%%%%%%%%%%%%%%%%%%%%%%
\begin{table}[]
\begin{tabular}{|p{40mm}|l|l|l|l|l|}
\hline
URANUS  &   $r_{min}$($\times 10^6 km$)          &   $r_{max}$($\times 10^6 km$)          &   $a$($\times 10^6 km$)          &    $b$($\times 10^6 km$)         &      eccentricity       \\ \hline \hline
 $N_{num}$       & 2729.595          & 2957.44 & 2843.519 & 2841.649275 & 0.0381 \\ \hline
  $N_{anal}$      & 2717.213 & 2984.63 & 2850.921 & 2847.766462 & 0.0469 \\ \hline
 $R_N=\frac{N_{num}}{N_{anal}}$\%        & 100.455  & 99.08 & 99.740 & 99.78519352 & 81.2504 \\ \hline
  $(N+YK)_{num}$      & 2729.595          & 2957.44 & 2843.519 & 2841.649245 & 0.0381 \\ \hline
  $(N+YK)_{anal}$      & 2717.213   & 2984.63 & 2850.921 &  2847.766434 & 0.0469 \\ \hline
 $R_{N+YK}$\\$=\frac{(N+YK)_{num}}{(N+YK)_{anal}}$\%        & 100.455 & 99.08 & 99.740 & 99.78519344 & 81.2504 \\ \hline
    Observation    & 2732.696          & 3001.39    &   2867.043      & \_\_\_\_\_  &  0.0469 \\ \hline
 $R^{N-Obs}_{num}$\\$= N_{num}/Obs$ \%    & 99.886        & 98.53 & 99.179 & \_\_\_\_\_  & 81.3684 \\ \hline
    $R^{N-Obs}_{anal}$\\$= N_{anal}/Obs$ \%      & 99.433  & 99.44 & 99.437 & \_\_\_\_\_  & 100.1452  \\ \hline
  $R^{YK-Obs}_{num}=$\\$  (N+YK)_{num}/Obs$  \%   & 99.886        & 98.53 & 99.179 & \_\_\_\_\_  &   81.3684 \\ \hline
   $R^{YK-Obs}_{anal}=$\\$ (N+YK)_{anal}/Obs$  \%         & 99.433 & 99.44  & 99.437 & \_\_\_\_\_  & 100.1452 \\ \hline
   \multicolumn{6}{|c|}{nominal $\a=10^{-8}$}\\ \hline
\end{tabular}
\caption{ The values of the calculated and observational astronomical parameters of the planet Uranus whose number of moons is 0} \label{TabAUra-N}
\end{table}
%%%%%%%%%%%%%%%%%%%%%%
\begin{table}[]
\begin{tabular}{|p{40mm}|l|l|l|l|l|}
\hline
URANUS  &   $r_{min}$($\times 10^6 km$)          &   $r_{max}$($\times 10^6 km$)          &   $a$($\times 10^6 km$)          &    $b$($\times 10^6 km$)         &      eccentricity       \\ \hline \hline
 $N_{num}$       & 2729.595          & 2957.44 & 2843.519 & 2841.649275 & 0.0381 \\ \hline
  $N_{anal}$      & 2717.213 & 2984.63 & 2850.921 & 2847.766462 & 0.0469 \\ \hline
 $R_N=\frac{N_{num}}{N_{anal}}$\%        & 100.455  & 99.08 & 99.740 & 99.78519352 & 81.2504 \\ \hline
  $(N+YK)_{num}$      & 2730.116          & 2992.91 & 2861.516 & 2858.935401 & 0.0441 \\ \hline
  $(N+YK)_{anal}$      & 2732.578   & 3001.50 & 2867.042 &  2863.869614 & 0.0469\\ \hline
 $R_{N+YK}$\\$=\frac{(N+YK)_{num}}{(N+YK)_{anal}}$\%        & 99.909 & 99.71 & 99.807 & 99.82770818 & 94.0932 \\ \hline
    Observation    & 2732.696          & 3001.39    &   2867.043      & \_\_\_\_\_  &  0.0469 \\ \hline
 $R^{N-Obs}_{num}$\\$= N_{num}/Obs$ \%    & 99.886        & 98.53 & 99.179 & \_\_\_\_\_  & 81.3684 \\ \hline
    $R^{N-Obs}_{anal}$\\$= N_{anal}/Obs$ \%      & 99.433  & 99.44 & 99.437 & \_\_\_\_\_  & 100.1452  \\ \hline
  $R^{YK-Obs}_{num}=$\\$  (N+YK)_{num}/Obs$  \%   & 99.905       & 99.71 & 99.807 & \_\_\_\_\_  &   94.2299   \\ \hline
   $R^{YK-Obs}_{anal}=$\\$ (N+YK)_{anal}/Obs$  \%         & 99.995 & 100.00 & 99.999 & \_\_\_\_\_  & 100.1452 \\ \hline
   \multicolumn{6}{|c|}{estimated $\a=-5.622864957252\times10^{-3}$}\\ \hline
   \end{tabular}
\caption{ The values of the calculated and observational astronomical parameters of the planet Uranus whose number of moons is 0} \label{TabAUra-E}
\end{table}
%%%%%%%%%%%%%%%%%%%%%
%%%%%%%%%%%%%%%%%%%%%
\begin{table}[]
\begin{tabular}{|p{40mm}|l|l|l|l|l|}
\hline
Neptune  &   $r_{min}$($\times 10^6 km$)          &   $r_{max}$($\times 10^6 km$)          &   $a$($\times 10^6 km$)          &    $b$($\times 10^6 km$)         &      eccentricity       \\ \hline \hline
 $N_{num}$       & 4464.81         & 4634.099 & 4549.454 & 4548.810665 & 0.0177 \\ \hline
  $N_{anal}$      & 4512.97 & 4601.381 & 4557.176 & 4556.953752 & 0.0097 \\ \hline
 $R_N=\frac{N_{num}}{N_{anal}}$\%        & 98.93  & 100.711 & 99.830 & 99.82130416 & 180.8744 \\ \hline
  $(N+YK)_{num}$      & 4464.81          & 4634.098 & 4549.454 & 4548.810617 & 0.0177 \\ \hline
  $(N+YK)_{anal}$      & 4512.97 & 4601.381 & 4557.176 &  4556.953706 & 0.0097 \\ \hline
 $R_{N+YK}$\\$=\frac{(N+YK)_{num}}{(N+YK)_{anal}}$\%        & 98.93 & 100.711 & 99.830  & 99.82130411 & 180.8743 \\ \hline
    Observation    & 4471.05          & 4558.857    & 4514.953      & \_\_\_\_\_  &  0.0097\\ \hline
 $R^{N-Obs}_{num}$\\$= N_{num}/Obs$ \%    & 99.86        & 101.650 & 100.764   & \_\_\_\_\_  & 182.6474 \\ \hline
    $R^{N-Obs}_{anal}$\\$= N_{anal}/Obs$ \%      & 100.93  & 100.932 & 100.935 & \_\_\_\_\_  & 100.9802  \\ \hline
  $R^{YK-Obs}_{num}=$\\$  (N+YK)_{num}/Obs$  \%   & 99.86       & 101.650 & 100.764 & \_\_\_\_\_  &   182.6473   \\ \hline
   $R^{YK-Obs}_{anal}=$\\$ (N+YK)_{anal}/Obs$  \%         & 100.93 & 100.932  & 100.935 & \_\_\_\_\_  & 100.9802 \\ \hline
   \multicolumn{6}{|c|}{nominal $\a=10^{-8}$}\\ \hline
\end{tabular}
\caption{ The values of the calculated and observational astronomical parameters of the planet Neptune whose number of moons is 0} \label{TabANep-N}
\end{table}
%%%%%%%%%%%%%%%%%%%
\begin{table}[]
\begin{tabular}{|p{40mm}|l|l|l|l|l|}
\hline
Neptune  &   $r_{min}$($\times 10^6 km$)          &   $r_{max}$($\times 10^6 km$)          &   $a$($\times 10^6 km$)          &    $b$($\times 10^6 km$)         &      eccentricity       \\ \hline \hline
 $N_{num}$       & 4464.81         & 4634.099 & 4549.454 & 4548.810665 & 0.0177 \\ \hline
  $N_{anal}$      & 4512.97 & 4601.381 & 4557.176 & 4556.953752 & 0.0097 \\ \hline
 $R_N=\frac{N_{num}}{N_{anal}}$\%        & 98.93  & 100.711 & 99.830 & 99.82130416 & 180.8744 \\ \hline
  $(N+YK)_{num}$      &4463.01          & 4546.479 & 4504.745 & 4504.517794 & 0.0096 \\ \hline
  $(N+YK)_{anal}$      & 4471.15 & 4558.747 & 4514.952   &  4514.73215 & 0.0097 \\ \hline
 $R_{N+YK}$\\$=\frac{(N+YK)_{num}}{(N+YK)_{anal}}$\%        & 99.81 & 99.730 & 99.773  & 99.77375499 & 98.6587 \\ \hline
    Observation    & 4471.05          & 4558.857    & 4514.953      & \_\_\_\_\_  &  0.0097\\ \hline
 $R^{N-Obs}_{num}$\\$= N_{num}/Obs$ \%    & 99.86        & 101.650 & 100.764   & \_\_\_\_\_  & 182.6474 \\ \hline
    $R^{N-Obs}_{anal}$\\$= N_{anal}/Obs$ \%      & 100.93  & 100.932 & 100.935 & \_\_\_\_\_  & 100.9802  \\ \hline
  $R^{YK-Obs}_{num}=$\\$  (N+YK)_{num}/Obs$  \%   & 99.82       & 99.728 & 99.773 & \_\_\_\_\_  &  99.6259   \\ \hline
   $R^{YK-Obs}_{anal}=$\\$ (N+YK)_{anal}/Obs$  \%         & 100.00 & 99.997  & 99.999 & \_\_\_\_\_  & 100.9802 \\ \hline
    \multicolumn{6}{|c|}{estimated $\a=9.351961741362\times10^{-3}$}\\ \hline
    \end{tabular}
\caption{ The values of the calculated and observational astronomical parameters of the planet Neptune whose number of moons is 0} \label{TabANep-E}
\end{table}
%%%%%%%%%%%%%%%%%%
%%%%%%%%%%%%%%%%
\begin{table}[]
\begin{tabular}{|p{40mm}|l|l|l|l|l|}
\hline
Pluto  &   $r_{min}$($\times 10^6 km$)          &   $r_{max}$($\times 10^6 km$)          &   $a$($\times 10^6 km$)          &    $b$($\times 10^6 km$)         &      eccentricity       \\ \hline \hline
 $N_{num}$       & 4439.709          & 7265.423 & 5852.566 & 5684.326067 & 0.2397 \\ \hline
  $N_{anal}$      & 4431.722 & 7298.614 & 5865.168 & 5687.267307 & 0.2444 \\ \hline
 $R_N=\frac{N_{num}}{N_{anal}}$\%        & 100.180  & 99.545 & 99.785 & 99.94828377 & 98.0832 \\ \hline
  $(N+YK)_{num}$      & 4439.709          & 7265.423 & 5852.566 & 5684.325992 & 0.2397 \\ \hline
  $(N+YK)_{anal}$      & 4431.722 & 7298.614 & 5865.168 &  5687.26725 & 0.2444 \\ \hline
 $R_{N+YK}$\\$=\frac{(N+YK)_{num}}{(N+YK)_{anal}}$\%        & 100.180 & 99.545 & 99.785 & 99.94828346 & 98.0832 \\ \hline
    Observation    & 4434.987          & 7304.326    & 5869.656      & \_\_\_\_\_  &  0.2444 \\ \hline
 $R^{N-Obs}_{num}$\\$= N_{num}/Obs$ \%    & 100.106       & 99.467 & 99.708 & \_\_\_\_\_  & 98.0882 \\ \hline
    $R^{N-Obs}_{anal}$\\$= N_{anal}/Obs$ \%      & 99.926  & 99.921 & 99.923 & \_\_\_\_\_  & 100.0051 \\ \hline
  $R^{YK-Obs}_{num}=$\\$  (N+YK)_{num}/Obs$  \%   &  100.106       & 99.467 & 99.708 & \_\_\_\_\_  &   98.0882 \\ \hline
   $R^{YK-Obs}_{anal}=$\\$ (N+YK)_{anal}/Obs$  \%         & 99.926 & 99.921  & 99.923 & \_\_\_\_\_  & 100.0051 \\ \hline
   \multicolumn{6}{|c|}{nominal $\a=10^{-8}$}\\ \hline
\end{tabular}
\caption{ The values of the calculated and observational astronomical parameters of the planet Pluto whose number of moons is 0} \label{TabAPlu-N}
\end{table}
%%%%%%%%%%%%%%%%%%%%%
\begin{table}[]
\begin{tabular}{|p{40mm}|l|l|l|l|l|}
\hline
Pluto  &   $r_{min}$($\times 10^6 km$)          &   $r_{max}$($\times 10^6 km$)          &   $a$($\times 10^6 km$)          &    $b$($\times 10^6 km$)         &      eccentricity       \\ \hline \hline
 $N_{num}$       & 4439.709          & 7265.423 & 5852.566 & 5684.326067 & 0.2397 \\ \hline
  $N_{anal}$      & 4431.722 & 7298.614 & 5865.168 & 5687.267307 & 0.2444 \\ \hline
 $R_N=\frac{N_{num}}{N_{anal}}$\%        & 100.180  & 99.545 & 99.785 & 99.94828377 & 98.0832 \\ \hline
  $(N+YK)_{num}$      & 4439.740          & 7280.242 & 5859.991 & 5690.112819 & 0.2407 \\ \hline
  $(N+YK)_{anal}$      & 4435.112 & 7304.196 & 5869.654 &  5691.616958 & 0.2444 \\ \hline
 $R_{N+YK}$\\$=\frac{(N+YK)_{num}}{(N+YK)_{anal}}$\%        & 100.104 & 99.672 & 99.835 & 99.97357273 & 98.4812 \\ \hline
    Observation    & 4434.987          & 7304.326    & 5869.656      & \_\_\_\_\_  &  0.2444 \\ \hline
 $R^{N-Obs}_{num}$\\$= N_{num}/Obs$ \%    & 100.106       & 99.467 & 99.708 & \_\_\_\_\_  & 98.0882 \\ \hline
    $R^{N-Obs}_{anal}$\\$= N_{anal}/Obs$ \%      & 99.926  & 99.921 & 99.923 & \_\_\_\_\_  & 100.0051 \\ \hline
  $R^{YK-Obs}_{num}=$\\$  (N+YK)_{num}/Obs$  \%   &  100.107       & 99.670 & 99.835 & \_\_\_\_\_  &   98.4862 \\ \hline
   $R^{YK-Obs}_{anal}=$\\$ (N+YK)_{anal}/Obs$  \%         & 100.002 & 99.998  & 99.999 & \_\_\_\_\_  & 100.0051 \\ \hline
   \multicolumn{6}{|c|}{estimated $\a=-7.642205983339201\times10^{-4}$}\\ \hline
\end{tabular}
\caption{ The values of the calculated and observational astronomical parameters of the planet Pluto whose number of moons is 0} \label{TabAPlu-E}
\end{table}

%%%%%%%%%%%%%%%%%%%%%%%%%
%%%%%%%%%%%%%%%%%%%%%%%%%%%
\clearpage
\appendix{\bf B. Tables of Absolute Deviations from Observation of the Planets}
\clearpage
\setcounter{table}{0}
\renewcommand{\thetable}{B\arabic{table}}

\begin{table}[]
\begin{tabular}{|l|l|l|l|l|l|} \hline
                      &     $R^{YK-Obs}_{num}$         & $R^{YK-Obs}_{anal}$             &    Observed $r_{max}$     &      $r^{num}_{max}-Obs$       &     $r^{anal}_{max}-Obs$             \\ \hline
\multicolumn{1}{|l|}{MERCURY} & 99.721 & 100.001  & 69.818   &  -0.194 & 0.001   \\  \hline
Venus                         & 99.769 & 100.018 & 108.941  & -0.251  & 0.020    \\ \hline
EARTH                         & 99.794 & 99.984 & 152.100    & -0.312  & -0.024 \\ \hline
MARS                          & 99.687 & 100.006 & 249.261  & -0.780  & 0.016    \\ \hline
JUPITER                       & 99.898  & 100.270 & 816.363  & -0.829 & 2.205   \\ \hline
SATURN                        & 101.116 & 100.794 & 1506.527 & 16.817 & 11.969   \\ \hline
URANUS                        & 98.535 & 99.441 & 3001.390  & -43.947  & -16.759    \\ \hline
Neptune                       & 101.650 & 100.932 & 4558.857 & 75.241   & 42.524    \\ \hline
Pluto                         & 99.467 & 99.921 & 7304.326 & -38.902  & -5.711  \\ \hline
\end{tabular}
\caption{ Absolute deviations, with nominal $\a$, of $r_{max}$ from observation, evaluated in ($10^6$ km). } \label{TabBrmax-N}
\end{table}
%%%%%%%%%%%%%%%%%%%%%%%%%
\begin{table}[]
\begin{tabular}{|l|l|l|l|l|l|} \hline
                      &     $R^{YK-Obs}_{num}$         & $R^{YK-Obs}_{anal}$             &    Observed $r_{min}$     &      $r^{num}_{min}-Obs$       &     $r^{anal}_{min}-Obs$             \\ \hline
\multicolumn{1}{|l|}{MERCURY} & 100.062 & 100.012  & 46.000   &  0.028 & 0.005   \\  \hline
Venus                         & 99.839 & 100.008 & 107.480  & -0.172  & 0.009   \\ \hline
EARTH                         & 99.857 & 99.978 & 147.095    & -0.210  & -0.031 \\ \hline
MARS                          & 99.962 & 99.999 & 206.650  & -0.077  & -0.001   \\ \hline
JUPITER                       & 99.906  & 100.262 & 740.595  & -0.692 & 1.947   \\ \hline
SATURN                        & 99.845 & 100.797 & 1357.554 & -2.092 & 10.824   \\ \hline
URANUS                        & 99.886 & 99.433 & 2732.696  & -3.100  & -15.482   \\ \hline
Neptune                       & 99.860 & 100.937 & 4471.050 & -6.239   & 41.922    \\ \hline
Pluto                         & 100.106 & 99.926 & 4434.987 & 4.722  & -3.264  \\ \hline
\end{tabular}
\caption{ Absolute deviations, with nominal $\a$, of $r_{min}$ from observation, evaluated in ($10^6$ km). } \label{TabBrmin-N}
\end{table}
%%%%%%%%%%%%%%%%%%%%%%%%%%%%%
\clearpage
\begin{table}[]
\begin{tabular}{|l|l|l|l|l|l|} \hline
                      &     $R^{YK-Obs}_{num}$         & $R^{YK-Obs}_{anal}$             &    Observed $r_{max}$     &      $r^{num}_{max}-Obs$       &     $r^{anal}_{max}-Obs$             \\ \hline
\multicolumn{1}{|l|}{MERCURY} & 99.706 & 99.995  & 69.818   &  -0.204 & -0.002   \\  \hline
Venus                         & 99.740 & 100.004 & 108.941  & -0.282  & 0.004    \\ \hline
EARTH                         & 99.757 & 99.997 & 152.100    & -0.369  & -0.003 \\ \hline
MARS                          & 99.664 & 99.996 & 249.261  & -0.835 & -0.008 \\ \hline
JUPITER                       & 99.334  & 100.003 & 816.363  & -5.430 & 0.027   \\ \hline
SATURN                        & 99.410 & 99.998 & 1506.527 & -8.874 & -0.019   \\ \hline
URANUS                        & 99.717 &  100.003 & 3001.390  & -8.472  & 0.117   \\ \hline
Neptune                       & 99.728 & 99.997 & 4558.857 & -12.377   & -0.109    \\ \hline
Pluto                         & 99.670 & 99.998 & 7304.326 & -24.083  & -0.12907  \\ \hline
\end{tabular}
\caption{ Absolute deviations, with estimated $\a$, of $r_{max}$ from observation, evaluated in ($10^6$ km). } \label{TabBrmax-E}
\end{table}
%%%%%%%%%%%%%%%%%%%%%%%%%
\begin{table}[]
\begin{tabular}{|l|l|l|l|l|l|} \hline
                      &     $R^{YK-Obs}_{num}$         & $R^{YK-Obs}_{anal}$             &    Observed $r_{min}$     &      $r^{num}_{min}-Obs$       &     $r^{anal}_{min}-Obs$             \\ \hline
\multicolumn{1}{|l|}{MERCURY} & 100.062 & 100.006  & 46.000   &  0.028 & 0.002   \\  \hline
Venus                         & 99.838 & 99.994 & 107.480  & -0.173  & -0.005   \\ \hline
EARTH                         & 99.856 & 100.003 & 147.095    & -0.211  & 0.004 \\ \hline
MARS                          & 99.962 & 99.989 & 206.650  & -0.077  & -0.022  \\ \hline
JUPITER                       & 99.897  & 99.996 & 740.595  & -0.757 & -0.027  \\ \hline
SATURN                        & 99.802 & 100.001 & 1357.554 & -2.684 & 0.020   \\ \hline
URANUS                        & 99.905 & 99.995 & 2732.696  & -2.579  & -0.117   \\ \hline
Neptune                       & 99.820 & 100.002 & 4471.050 & -8.037  & 0.107    \\ \hline
Pluto                         & 100.107 & 100.002 & 4434.987 & 4.753  & 0.125  \\ \hline
\end{tabular}
\caption{ Absolute deviations, with estimated $\a$, of $r_{min}$ from observation, evaluated in ($10^6$ km). } \label{TabBrmin-E}
\end{table}
%%%%%%%%%%%%%%%%%%%%%%%
%%%%%%%%%%%%%%%%%%%%%%%%%%

\end{appendices}


\clearpage
\begin{thebibliography}{999}
% Reference 1

\bibitem{0}
  Ephraim Fischbach and Carrick L. Talmadge, The Search for Non-Newtonian Gravity, AIP Press, Springer (1999),


\bibitem{1}	L. D. Landau and E. M. Lifshitz, Course of Theoretical Physics (Mechanics), Vols. 1 Ch 3, Sec 14, p 32, (Pergamon Press : Oxford), (1969).



\bibitem{2} 	I. Rodriguez and J. L. Brun, "Closed orbits in central forces distinct from Coulomb or harmonic oscillator type," European Journal of Physics, vol. 19, pp. 41-49, 1998.

\bibitem{3}	. J. L. Brun and A. F. Pacheco, "On closed but non-geometrically similar orbits," Celestial Mech Dyn Astr, pp. 311-316, 2006.

 \bibitem{Hinterbichler}
K. Hinterbichler, ``Theoretical Aspects of Massive Gravity'', Rev. Mod. Phys. 84, 671 (2012), arXiv:1105.3735 [hep-th].

\bibitem{deRham}
Claudia de Rham, ``Massive Gravity'', Living Reviews in Relativity, 17 (2014) 7, arXiv:1401.4173 [hep-th].

\bibitem{Goldhaber}
A. S. Goldhabert and M. M. Nieto, ``Mass of the graviton'', PRD9, 1119 (1974)



\bibitem{Dong}
    Yiming Dong, Lijing Shao, Zexin Hu, Xueli Miaoc and
Ziming Wang, ``Prospects for Constraining the
Yukawa Gravity with Pulsars
around Sagittarius A*", JCAP-11-(2022)-051, arXiv:2210.16130 [astro-ph.HE].



\bibitem{4}	J. W. Moffat, "Scalar-tensor-vector gravity theory," Journal of Cosmology and Astroparticle Physics, 2006.

\bibitem{Zhang}
Xing Zhang, Tan Liu and Wen Zhao, ``Gravitational radiation from compact binary systems in screened modified gravity", Phys. Rev.D95, 104027 (2017), arXiv: 1702.08752 [gr-qc].

\bibitem{DAddio}
A. D’Addio, R. Casadio, A. Giusti, and M. De Laurentis, ``Orbits in bootstrapped Newtonian gravity'', Phys. Rev. D 105, 104010
(2021), arXiv:2110.08379 [gr-qc]

\bibitem{Monica}
R. Della Monica, I. de Martino, M. De Laurentis, ``Orbital precession of the S2 star in Scalar–Tensor–Vector Gravity'', Monthly Notices of the Royal Astronomical Society, Volume 510, Issue 4, March 2022, Pages 4757–4766,
arXiv:2105.12687 [gr-qc]

\bibitem{DAMTP}
David Benisty, ``Testing modified gravity via Yukawa potential in two body problem: Analytical
solution and observational constraints'', Phys. Rev. D106, 043001 (2022).


\bibitem{Banik} I. Banik and H. Zhao, Mon. Not. Roy. Astron. Soc. 480,
2660 (2018), [Erratum: Mon.Not.Roy.Astron.Soc. 482,
3453 (2019), Erratum: Mon.Not.Roy.Astron.Soc. 484,
1589 (2019)], arXiv:1805.12273 [astro-ph.GA].


\bibitem{Lu} Q. Yu, F. Zhang, and Y. Lu, Astrophys. J. 827, 114
(2016), arXiv:1606.07725 [astro-ph.HE].

\bibitem{Pricopi} D. Pricopi, Astrophys. Space Sci. 361, 277 (2016).

\bibitem{Edwards} J. P. Edwards, U. Gerber, C. Schubert, M. A. Trejo, and
A.Weber, PTEP 2017, 083A01 (2017), arXiv:1706.09979
[physics.atom-ph].

\bibitem{Mukherjee} R. Mukherjee and S. Sounda, Indian Journal of Physics
92, 197 (2018), arXiv:1705.02444 [physics.plasm-ph].

\bibitem{Iorio}
Lorenzo Iorio, ``Putting Yukawa-LikeModified Gravity (MOG) on
the Test in the Solar System",  Scholarly Research Exchange,
Volume 2008, Article ID 238385

\bibitem{Laurentis}
M. De Laurentis, I. De Martino, and R. Lazkoz, ``Analysis of the Yukawa gravitational potential in f(R) gravity II:
relativistic periastron advance'',  Phys.
Rev. D 97, 104068 (2018), arXiv:1801.08136 [gr-qc].

\bibitem{Berge}
J. Bergé, P. Brax, M. Pernot-Borràs, and J.-P. Uzan, ``Interpretation of geodesy experiments in
non-Newtonian theories of gravity",
Class. Quant. Grav. 35, 234001 (2018), arXiv:1808.00340
[gr-qc].

\bibitem{Lorenzo}
Lorenzo Iorio, ``Constraints on a Yukawa gravitational potential
from laser data of LAGEOS satellites'', Physics Letters A 298 (2002) 315–318










\bibitem{5}	E. Cavan, I. Haranas, I. Gkigkitzis and K. Cobbett, "Dynamics and stability of the two body problem with Yukawa correction," Astrophysics and Space Science, 2020.
\bibitem{6} 	"https://nssdc.gsfc.nasa.gov/planetary/factsheet/index.html," [Online].
\bibitem{7} 	A. Rujula, dedicated to viktor weisskopf on the occasion of the viki-fest, erice, 1986.
\bibitem{8}	H. Goldstein, Classical Mechanics, SECOND EDITION ed., ADDISON-WESLEY PUBLISHING COMPANY , 1980.
\bibitem{9}	J. D. Meiss, Differential Dynamical Systems, 2007.
\bibitem{10}	O. FACKLER and J. T. T. VAN, 5th FORCE NEUTRINO PHYSICS, 1988.
\bibitem{11}	S. L. Ross, Differential Equation (John Willey \& Sons), 1984, p. 661.
\bibitem{12}	K. Wakker, Fundamentals of Astrodynamics, 2015.




\end{thebibliography}
\end{document}